\documentclass[twocolumn]{aastex7}
\usepackage{amsmath,amssymb,gensymb,times,graphicx,morefloats,color,wrapfig}

\newcommand{\tess}{\textit{TESS}}
\newcommand{\prot}{$P_{\rm rot}$}
\newcommand{\ktwo}{{\textit K2}}
\newcommand{\kepler}{{\it Kepler}}
\newcommand{\gaia}{\textit{Gaia}}
\newcommand{\teff}{$T_{\rm eff}$}
\newcommand{\kms}{\ensuremath{\mathrm{km~s^{-1}}}}
\newcommand{\complex}{Greater Pleiades Complex}
\newcommand{\upperright}{GPC-1}
\newcommand{\lowerright}{GPC-2}
\shorttitle{The Dissolving Pleiades Complex}
\shortauthors{Boyle et al.}

\begin{document}


\title{Lost Sisters Found: \tess\ and \gaia\ Reveal a Dissolving Pleiades Complex}

\author[orcid=0000-0001-6037-2971]{Andrew W. Boyle}
\affiliation{Department of Physics and Astronomy, The University of North Carolina at Chapel Hill, Chapel Hill, NC 27599, USA}
\altaffiliation{NSF Graduate Research Fellow}
\email[show]{awboyle@unc.edu}  

\author[orcid=0000-0002-0514-5538]{Luke G. Bouma}
\affiliation{Observatories of the Carnegie Institution for Science, Pasadena, CA 91101, USA}
\altaffiliation{Carnegie Fellow}
\email{lbouma@carnegiescience.edu}

\author[orcid=0000-0003-3654-1602]{Andrew W. Mann}
\affiliation{Department of Physics and Astronomy, The University of North Carolina at Chapel Hill, Chapel Hill, NC 27599, USA}
\email{awmann@unc.edu}

\correspondingauthor{Andrew Boyle}

\begin{abstract}



Most star clusters dissolve into the Galaxy over tens to hundreds of millions of years after they form.
While recent \gaia\ studies have honed our view of cluster dispersal, the exact chronology of which star formation events begat which star cluster remnants remains unclear.  This problem is acute after 100~Myr, when cluster remnants have spread over hundreds of parsecs and most age estimates for main sequence stars are too imprecise to link the stars to their birth events.
Here we develop a Bayesian framework that combines \tess\ stellar rotation rates with \gaia\ kinematics to identify diffuse remnants of open clusters. We apply our method to the Pleiades, which previous studies have
noted shows kinematic similarities to other nearby young stellar groups. We find that the Pleiades constitutes the bound core of a much larger, coeval structure that contains multiple known clusters distributed over 600~pc.  We refer to this structure as the Greater Pleiades Complex.  On the basis of uniform ages, coherent space velocities, detailed elemental abundances, and traceback histories, we conclude that most stars in this complex originated from the same giant molecular cloud. This work establishes a scalable approach for tracing the genealogies of nearby clusters and further cements the Pleiades as a cornerstone of stellar astrophysics. We aim to apply this methodology to other associations as part of the upcoming \tess\ All-Sky Rotation Survey.



\end{abstract}



\section{Introduction} 

Clusters of stars that formed from the same collapsing molecular cloud are powerful tools for a wide range of astrophysics. In stellar physics, star clusters provide stringent tests on the evolution of stellar structure \citep{Stassun2014,Feiden2016}, stellar rotation \citep{Curtis2019,Rebull2020}, and stellar activity \citep{Agueros2018,Kiman2021}. In exoplanet science, young stars with well-measured ages are high-priority targets for direct imaging \citep[][]{Bowler2016,Nielsen2019} and transiting \citep[][]{Newton2019, Barber2024} exoplanet searches, as they provide unique insight into the formation and early evolution of planetary systems \citep[][]{Vach2024,Thao2024a,Fernandes2025}.  In galactic and extragalactic contexts, young clusters reveal the local star formation history and are also used to study spiral arms, tidal forces, and molecular clouds \citep[][]{Krumholz2019,Cantat-Gaudin2024}. 

However, the vast majority of star clusters dissolve shortly after their formation. Within ten megayears following cloud collapse, the star-forming gas is ejected by jets, winds, radiation, and supernovae \citep{Lada2003,Krumholz2019,Grudic2022}.  Embedded clusters thus become open, and most of their stars become gravitationally unbound.  The small subset of clusters which remain bound rarely survive the next gigayear due to internal forces \citep{Baumgardt2007, PortegiesZwart2010}, collisions with giant molecular clouds \citep{Gieles2006, Miller2025}, and perturbations by the Galactic tide \citep{Gieles2008}.  The dissolution of clusters---whether due to gas expulsion or later evolution---gives rise to unbound stellar associations which can remain spatially and kinematically coherent for hundreds of millions of years.

The hierarchical structure of currently intact young associations (e.g., Orion, Sco-Cen, Vela, h/$\chi$ Per) implies that some older, already known cluster remnants may be dissolved remnants of individual star formation events. Indeed, recent studies have highlighted that many of the nearby moving groups and associations share similar kinematic properties \citep[e.g.,][]{gagneNumberNearbyMoving2021, Kerr2022}. Current star formation near the Sun similarly seems to be driven by expansion of the local bubble \citep{Zucker2022}.

Astrometry from \gaia{} has revolutionized our ability to identify members of dissolving associations. Unsupervised clustering algorithms in particular have revealed hundreds of previously unknown groups \citep[e.g.,][]{kounkelUntanglingGalaxyII2020,Moranta2022,Hunt2023}. However, since these techniques rely on the density of stars in the phase space of positions and velocities, they struggle to detect the most dispersed remnants. Once stars are sufficiently far from their birthplaces, they may no longer be detectable with purely spatial or kinematic clustering. Some efforts improve the selection by restricting the sample using other information;  \citet{kerrStarsPhotometricallyYoung2021}, for example, limited their initial search to pre-main sequence stars using photometric selection.  This increases the contrast of bona fide cluster stars relative to the field.

Stellar rotation provides an alternative means for selecting young stars and hence detecting weaker overdensities than possible with the Gaia data alone. Stellar rotation periods increase with age due to magnetic braking, a relationship known as gyrochronology \citep{skumanichTimeScalesCA1972,barnesAgesIllustrativeField2007,Bouma2023}. Rapid rotation can thus also serve as a powerful discriminator between young stars and the field; the only major interlopers are binaries \citep{Simonian2019}, which are often easy to identify \citep{Wood2021}. \tess\ has now surveyed over 90\% of the sky, which dramatically increases the potential for rotation-based selection of young stellar populations. 

In order to apply rotation-based selection reliably, we need to understand the precision and accuracy of stellar rotation periods from \tess\ photometry, as well as how to interpret non- or weak-detections. In \citet{2025arXiv250413262B}, henceforth B25, we explored the uncertainty, reliability, and precision of \prot\ estimates derived using \tess\ light curves corrected using a Causal Pixel Model \citep{wangCausalDatadrivenApproach2016, hattoriUnpopularPackageDatadriven2021} and applying a Lomb-Scargle periodogram \citep{lombLeastsquaresFrequencyAnalysis1976}. One of our goals was to help search for new young associations and members of known associations.

Here, we combine the method of B25 and kinematics from \gaia\ to develop a rotation-based Bayesian membership calculation (gyro-tagging) capable of identifying the dissolving and diffuse parts of open clusters. As a proof of concept, we aim to see if we can recover any extensions to the Pleiades, which has previously been noted to be kinematically similar to certain nearby associations \citep{gagneNumberNearbyMoving2021}. Our method uncovers the \textit{Greater Pleiades Complex}---a sprawling, kinematically connected structure that likely originated from the same star formation event as the Pleiades and that harbors a wide range of known groups, including the nearby young moving group AB Dor. Our results confirm that we can uncover large, diffuse associations and lays the groundwork for applying our methodology to search for other dissolving associations in the upcoming \tess\ All-Sky Rotation Survey main paper (Boyle et al., in prep.). 

This paper is structured as follows. In Section \ref{sec:survey}, we discuss the target selection, reddening correction, and temperature calculation for all stars in the \tess\ All-Sky Rotation Survey. We select candidate Pleiades Complex members in Secion~\ref{sec:pleiades} and assign membership probabilities to each of these stars in Section~\ref{sec:bayes}. We then use kinematic back integrations, chemical abundances, and multiple youth indicators to analyze the subgroups of the complex. We discuss our results in Section~\ref{sec:discussion} and conclude in Section~\ref{sec:conclusion}.

\section{The \tess\ All-Sky Rotation Survey} \label{sec:survey}

NASA's \tess\ mission \citep{Ricker2015} provides time-series observations for millions of stars across the entire sky, enabling rotation measurements on an unprecedented scale. 
The \tess\ All-Sky Rotation Survey is a program aiming to estimate stellar rotation periods for every fast-rotating star $T < 16$ star within 500 pc. The resulting catalog opens new avenues to explore the age and clustering of nearby stars through their rotational behavior. The primary goal of the \tess\ All-Sky Rotation Survey is to identify new, dissolving stellar associations, expand membership lists for already known associations, and search for bridges between known associations that could indicate a common origin.
We describe the construction of this catalog below. 

\subsection{\gaia\ data}

We create an initial list of nearby stars by selecting all stars from the \tess\ Input Catalog v8.2 \citep[TIC v8.2;][]{stassunRevisedTESSInput2019} with $T < 16$ and $d < 500$ pc. This results in 7,883,661 stars, of which 127,202 are flagged as duplicate entries, which we dropped. We then cross-matched these stars with \gaia\ DR3, taking the closest match within a 1\arcsec{} from the TIC coordinates. For all stars with no match, we used the \gaia\ DR3 \texttt{dr2\_neighbourhood} table to find stars within 2\arcsec{} of the \gaia\ DR2 identifier given in the TIC. This resulted in 7,828,017 matches. 

We ran each of these targets through the \texttt{tess-point} tool \citep{2020ascl.soft03001B} to identify which stars fall on \tess\ silicon between 2018 July and 2026 September
(Sectors 1-107). This resulted in 6,873,227 stars that will have at least one sector of \tess\ data by the end of the currently-approved \tess\ extended missions. Data for Sectors 1--94 was available at the time of this analysis.

\begin{figure*}
    \centering
    \includegraphics[width=1\linewidth]{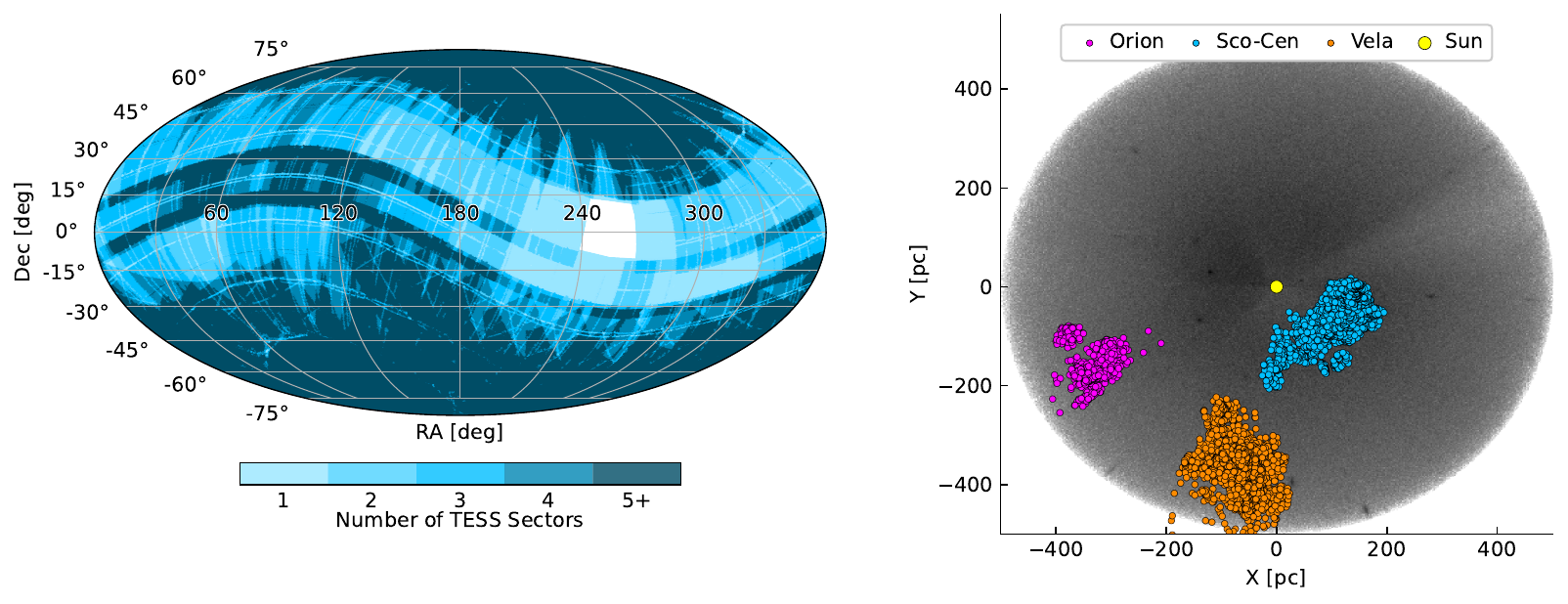}
    \caption{\textbf{\tess's near all-sky coverage enables studies of large stellar complexes.} \textit{Left:} \tess\ coverage from 2018 July to 2026 August (Sectors 1-107), color coded by the number of months over which each star has been observed by \tess. \textit{Right:} Galactocentric X and Y positions of Orion ($t<12$ Myr), Sco-Cen ($t\sim17$ Myr), and Vela ($t\sim19$ Myr), with the Sun at the origin. Even at young ages, these hierarchically-structured groups span hundreds of parsecs in X and Y. Gray points are all stars in our $T < 16$, $d < 500$ pc target list. The membership list for Orion is taken from \citet{2018AJ....156...84K} (excluding stars in that work marked as field stars or not assigned to an Orion sub-group), Sco-Cen from \citet{2023ApJ...954..134K}, and Vela from \citep{2019A&A...626A..17C} (including only stars with a $>99$\% probability from that work of being part of a Vela sub-group). Positive Y is in the direction of Galactic rotation and positive X is in the direction of the Galactic center. This will be the orientation of our plots for the rest of this work.}
    \label{fig:fig1}
\end{figure*}

Figure~\ref{fig:fig1} shows why \tess\ is a powerful tool for this type of study: its near all-sky coverage allows for detailed studies of nearby stellar complexes, even when those complexes are spread over hundreds of parsecs.

\subsection{Extinction and Effective Temperature Calculation}\label{subsec:reddening}

Gyrochronology relations are typically calibrated as a function of effective temperatures (\teff) or intrinsic color. Extinction will be significant for stars beyond $\simeq$200\,pc (most of our sample), so we corrected for extinction and derived \teff{} from \gaia\ DR3 \( G_{\rm BP} - G_{\rm RP} \) colors in a homogeneous way. 

For extinction corrections, we used the three-dimensional dust maps from \citet{2022A&A...664A.174V}, which were constructed using a hierarchical Bayesian approach that incorporates \gaia, 2MASS, and Pan-STARRS photometry. These maps span a \(3000\,\mathrm{pc} \times 3000\,\mathrm{pc} \times 800\,\mathrm{pc}\) volume around the Sun with a spatial resolution of 10\,pc. 

For each star, we integrated extinction along the line of sight to its \gaia-inferred distance using the \citet{2022A&A...664A.174V} cube, which reports extinction in units of \( A_0 \) per parsec on a Cartesian grid centered on the Sun. We then converted the cumulative \( A_0 \) into band-specific extinctions (\( A_G \), \( A_{\rm BP} \), \( A_{\rm RP} \)) using the extinction law from \citet{2019ApJ...886..108F}, calibrated for Gaia EDR3. Since the Gaia DR3 extinction coefficients\footnote{\url{https://www.cosmos.esa.int/web/gaia/edr3-extinction-law}} depend on the de-extincted \( G_{\rm BP} - G_{\rm RP} \) color, we performed the correction iteratively until it changed by less than 0.01\,mag. We applied the final corrections to all three \gaia\ bands. 

We computed effective temperatures following the procedure described by \citet{curtisWhenStalledStars2020}. That work established an empirical relation between spectroscopic temperatures and Gaia DR2 \( G_{\rm BP} - G_{\rm RP} \) colors. Here, we update the analysis to use our dereddened Gaia DR3 colors and apply the resulting empirical relation to estimate \teff\ for each star in our catalog. 

Gyrochronology applies only to stars whose rotation is slowed by magnetic braking due to stellar winds. Effective magnetic braking requires stars to have substantial convective envelopes, which sustain the magnetic dynamo responsible for threading the stellar wind with field lines \citep{2014ARA&A..52..251C}. Above a stellar mass of ${\sim}1.5\,M_\odot$, the outer convection zone vanishes, halting the dynamo and rendering magnetic braking ineffective \citep{2015nps..book.....I}. This transition, known as the Kraft break, occurs at an effective temperature of ${T_{\rm eff}\simeq 6550\,{\rm K}}$ \citep{kraftStudiesStellarRotation1967,2024ApJ...973...28B}. At the opposite extreme, fully convective M dwarfs also deviate from standard gyrochronology relations \citep{2024NatAs...8..223L}, and their intrinsic faintness complicates rotation period measurements. To ensure gyrochronology validity and observational feasibility, we restrict our sample to stars with $3000\,{\rm K} < T_{\rm eff} < 6500\,{\rm K}$, removing stars lacking \textit{Gaia} $G$, $G_{\rm BP}$, or $G_{\rm RP}$ photometry. This yields a final target list of 6,662,818 stars used for all subsequent analysis.



\subsection{Rotation periods from \tess} \label{subsec:rotation_measurement}

For each star in our membership list, we generated \tess\ light curves using the \texttt{unpopular} package \citep{hattoriUnpopularPackageDatadriven2021}, with the parameters from B25. Briefly, \texttt{unpopular} uses causal pixel modeling (CPM) to isolate intrinsic stellar variability, such as rotation, while mitigating instrumental systematics.

For each sector in which a star was observed, we computed a Lomb-Scargle periodogram on a uniform frequency grid from 0.2--20 days. We adopted the period corresponding to the highest Lomb-Scargle power across all sectors. We consider all rotation period measurements with a Lomb-Scargle power $>0.05$ as being a detection and all others as a non-detection. Based on the reliability and completeness mapping of B25, this threshold removes about 30\% of stars with \prot\ $<12$ days from our sample, but also increases the reliability of the periods we do measure by $\sim$60\%. Because the groups in question are diffuse, there are many more field stars than members in any given region; hence reliability is more important (removing false positives) than completeness (catching all fast rotators). 




\section{The Pleiades Complex} \label{sec:pleiades}

The Pleiades (Melotte 22; Subaru; the Seven Sisters) is one of the nearest (136\,pc) and most prominent open star clusters, and it has been a cornerstone of stellar astrophysics for over a century. Its age \citep[$\sim$127 Myr;][]{2022A&A...664A..70G} and mass \citep[$\sim850M_{\odot}$;][]{meingastExtendedStellarSystems2019a}, make it an important region to study stellar spin-down \citep{rebullROTATIONPLEIADESK22016}, the stellar initial mass function \citep{Lodieu2012}, and search for young planets \citep{Gaidos2017}.

Recently, \cite{gagneNumberNearbyMoving2021} hypothesized that the Pleiades and several nearby moving groups might be related due to their similar kinematics and age. They noted that AB Doradus, Theia 234, Theia 301, Theia 368, and Theia 369 all move within $5 \rm ~km~s^{-1}$ of each other in UVW space. 
Given that one of the major goals of the \tess\ All-Sky Rotation Survey is to identify the dissolving and diffuse parts of open clusters, we chose to use the Pleiades as a test case for our methodology to see if we could confirm the connection between the Pleiades and any of its hypothesized extensions. 

\subsection{Initial Target Selection} \label{subsec:target_selection}

The hallmark properties of open clusters are a shared age, chemical composition, and coherent distributions in both position and velocity. Accordingly, most clustering analyses to identify open clusters are performed in five dimensions (three spatial and two velocity components; e.g., \citealt{kounkelUntanglingGalaxyLocal2019, kerrStarsPhotometricallyYoung2021, Hunt2023}) or in six dimensions when sufficient radial velocities are available (e.g., \citealt{Moranta2022}). Clustering in all six phase-space dimensions suppresses field star contamination better than clustering in five dimensions \citep{Moranta2022, Boyle2023}. However, six-dimensional analyses are limited by the availability of radial velocity measurements: Gaia DR3 provides RVs for only $\sim$1.9\% of the full catalog of 1.8 billion sources \citep{2023A&A...674A...5K} and is limited to stars with $G_{\rm RVS} < 14$. \gaia\ radial velocities are also less precise than the tangential velocities derived from \gaia\ proper motions, especially for stars within 500\,pc. As a result, clustering exclusively in the traditional six dimensions sacrifices completeness. 

Here, we use the \tess\ all-sky rotation sample to fill the gap. Specifically, we perform a six-dimensional clustering analysis to identify a high-confidence sample and delineate the Pleiades' structure (Section~\ref{subsec:hdbscan}). We then expand this sample by including stars lacking radial velocities via a complementary five-dimensional comoving selection (Section~\ref{subsec:comove}).

A summary of our process for identification and construction of the \complex{} is:
\begin{enumerate}
    \item Select \gaia\ stars with three-dimensional velocity $<$5 \kms{} of the Pleiades Core.
    \item Select stars from (1) that have \tess\ rotation periods less than 12 days. 
    \item Apply \added{HDBSCAN} to the sample from (2) (Section~\ref{subsec:hdbscan}).
    \item Remove any stars from (3) that have rotation periods or CMD positions that are inconsistent with the Pleiades core. This defines the central (high-likelihood) members.
    \item Add in stars with tangential two-dimensional velocities matching central members using \texttt{Comove} (Section~\ref{subsec:comove}).
    \item Calculate Bayesian membership probabilities including rotation (Section~\ref{sec:bayes}).
    \item Assess association with Pleiades using age and abundance estimates and tracing groups back kinematically (Section~\ref{sec:results}).
\end{enumerate}


To define a set of core Pleiades members, we selected Pleiads from \citet{Hunt2023} with a reported membership probability above 0.9. After excluding stars lacking radial velocity measurements in Gaia DR3, we transformed the Gaia astrometry and radial velocities into Galactocentric UVW velocities and adopted the median UVW of the remaining stars as the cluster's core velocity. We used the \texttt{v4.0} Galactocentric coordinate frame in \texttt{astropy}, which comes with the Sun's distance to the galactic center of 8,122 pc, solar velocity of (U, V, W)$_{\odot}$ = (12.9, 245.6, 7.78) km~s$^{-1}$, and distance above the galactic plane of 20.8 pc.

Next, we selected all stars from the list defined in Section \ref{sec:survey} that lie within $5$ \kms{} of the Pleiades' median velocity. This search radius matches the velocity threshold used by \citet{gagneNumberNearbyMoving2021} to connect sub-populations. Numerical simulations similarly suggest that dissolving populations can have velocity dispersions $\gtrsim5$\kms{} after 100 Myr \citep[e.g.,][]{jerabkova800PcLong2021,Boyle2023}.  Empirical studies of other young associations similarly report internal dispersions spanning $0.5$--$5~\mathrm{km~s^{-1}}$ \citep{ 2018ApJ...862..138G,Kuhn2019,2021A&A...647A..91G}. Lastly, any search bound needs to account not just for internal velocity spread, but spread due to systematic and measurement uncertainties in the velocities themselves, especially in the RVs. 

This selection yielded 10,234 candidate members. We calculated rotation periods for each as described in Section~\ref{subsec:rotation_measurement} and assessed the reliability of each period using the probabilistic framework introduced in B25. This framework returns the probability that a measured period is correct, conditional on its value and associated Lomb-Scargle power. We retained only those periods with reliability $>$0.7.

Finally, prior studies of 120 Myr-old populations have shown that stars with rotation periods $>$$12$ days are rare at this age \citep{rebullROTATIONPLEIADESDATA2016, curtisTESSRevealsThat2019, 2020MNRAS.492.1008G}, so we removed such long-period detections. After these quality cuts, our working sample contained 3,489 stars.

\subsection{Initial Clustering with HDBSCAN} \label{subsec:hdbscan}

\begin{figure*}
    \centering
    \includegraphics[width=1\linewidth]{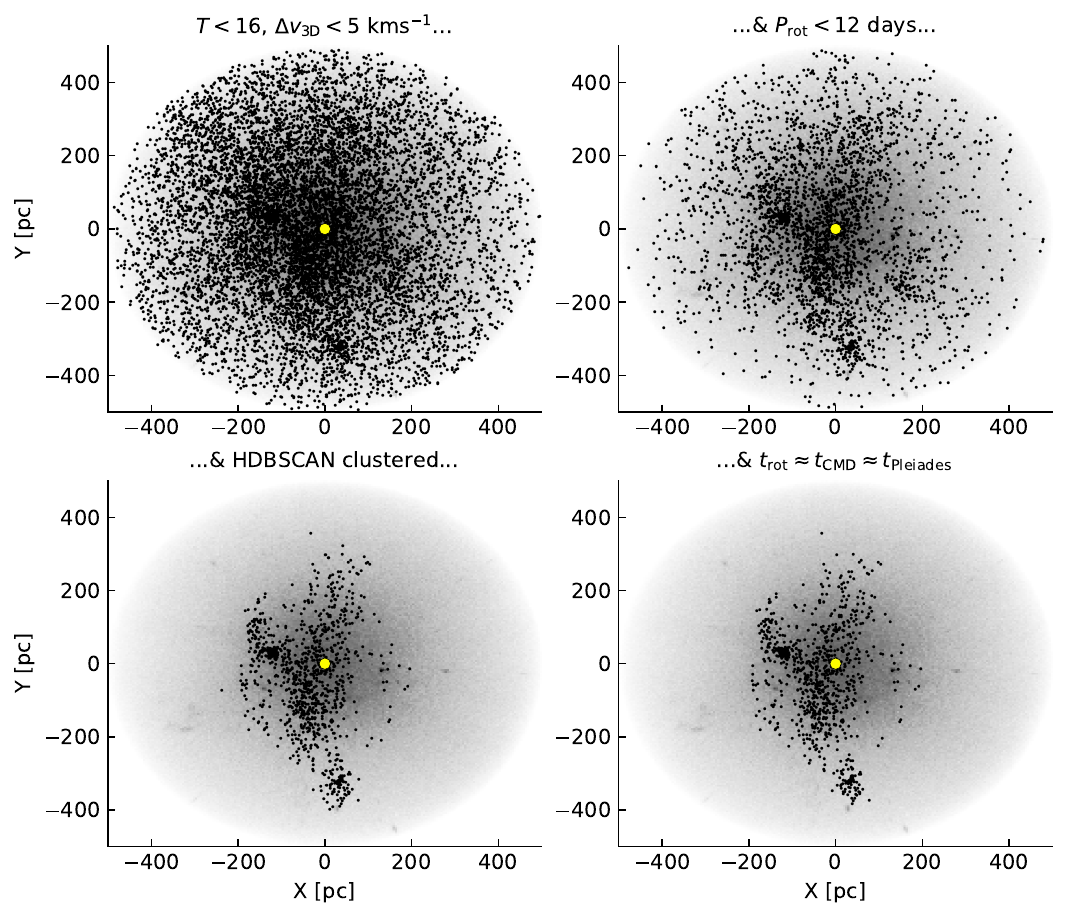}
    \caption{\textbf{Stellar kinematics and rotation isolate the \complex}. \textit{Top Left}: Stars with $T < 16$ and $d < 500$ pc (gray), highlighting those within $5~\mathrm{km~s^{-1}}$ of the Pleiades' median 3D velocity (black). The Sun is marked by a yellow point at the origin. \textit{Top right}: Subset of stars consistent with the Pleiades in velocity, with reliable \tess\ rotation periods (\prot\ $<12$ days and reliability $>0.7$). \textit{Bottom left}: Six-dimensional clustering results from HDBSCAN applied to the velocity and rotation-selected sample. \textit{Bottom right}: High-confidence members ($N$=1,075) with 3D velocities, selected for spatiokinematic coherence, rotation consistency, and placement on the cluster CMD. These stars define a co-moving and coeval population that we will argue comprises the Greater Pleiades Complex.}
    \label{fig:quality_cuts}
\end{figure*}

The top-right panel of Figure~\ref{fig:quality_cuts} shows the target list after the quality cuts in the previous section: beyond the Pleiades there are many stars scattered over hundreds of parsecs. To identify clustered populations and remove outliers, we applied the HDBSCAN clustering algorithm \citep{mcinnesHdbscanHierarchicalDensity2017}. HDBSCAN identifies clusters based on local point density and does not require prior assumptions about cluster number or shape, making it well-suited to uncover both compact cores and diffuse halos. For the stellar input, we used stellar positions and velocities in Galactocentric Cartesian coordinates (XYZUVW), and normalized each dimension to zero mean and unit variance.  Other choices for normalization do exist \citep{kerrStarsPhotometricallyYoung2021,Hunt2023}, and would likely yield subtly different results. 


Our clustering setup used three key hyperparameters. The first, \texttt{min\_cluster\_size}, defines the minimum number of stars required for a group to be considered a distinct cluster. We adopted \texttt{min\_cluster\_size} = 10 to allow sensitivity to small-scale substructure while suppressing spurious groupings. The second parameter, \texttt{min\_samples}, influences the algorithm’s conservativeness: higher values tend to produce more robust (but less complete) clusters by labeling more stars as noise. We set \texttt{min\_samples} = 10, matching the cluster size threshold. The third parameter, \texttt{cluster\_selection\_epsilon}, controls how HDBSCAN handles regions of varying density—especially useful for resolving clusters connected by diffuse bridges of stars. Without this setting, the algorithm can artificially fragment coherent structures. We also used the ``Excess of Mass'' (EOM) mode to avoid splitting structures. We tested a range of reasonable parameter values and found our results to be broadly stable, provided \texttt{cluster\_selection\_epsilon} was defined, so used \texttt{cluster\_selection\_epsilon} = 1. In all configurations, HDBSCAN revealed an extended structure surrounding the dense cluster core. To obtain a high-confidence sample for further analysis, we retained only stars with HDBSCAN membership probabilities $>0.5$. This selection yielded 1,254 stars. For completeness, we also repeated this analysis using a $10~\mathrm{km~s^{-1}}$ velocity cut instead of the $5~\mathrm{km~s^{-1}}$ velocity cut and found no significant additions to the structure.

\begin{figure*}
    \centering
    \includegraphics[width=0.95\linewidth]{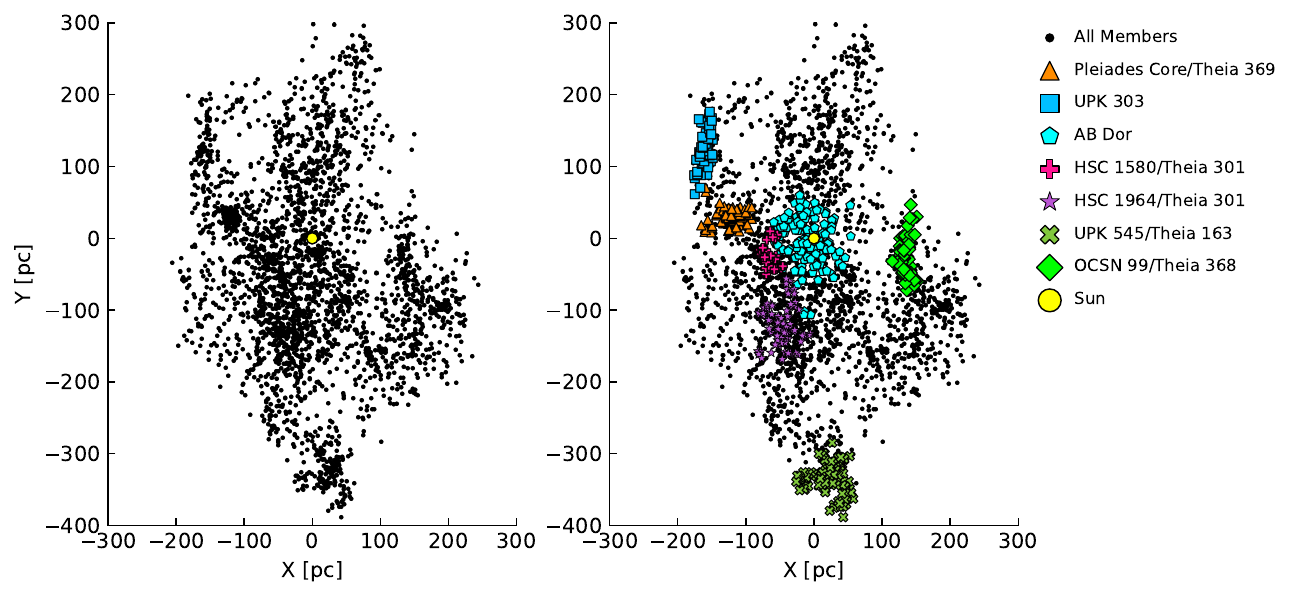}
    \caption{\textbf{The \complex{} connects multiple known associations}. \textit{Left:} An XY projection of the stars in our final membership list with a probability $P\bigl(M \mid P_{\rm rot}\bigr) > 0.5$. This includes stars without radial velocity measurements that were not present in Figure~\ref{fig:quality_cuts}. \textit{Right}: The results of cross-matching our membership list with the groups from \cite{kounkelUntanglingGalaxyLocal2019} and \cite{Hunt2023}, keeping only matches with more than 50 stars. Stars from AB Dor are taken from the Montreal Open Clusters and Associations (MOCA) database (J. Gagn\'e et al., in preparation; \citealt{2024PASP..136f3001G}). Multiple known groups in the region of the Pleiades appear to be connected by a bridge of coeval, comoving stars.}
    \label{fig:groups}
\end{figure*}

To further refine our membership list, we used stellar rotation and photometry as age diagnostics. We first compared our candidate members against Pleiads from \citet{rebullROTATIONPLEIADESDATA2016} in the $P_{\rm rot}$ versus $T_{\rm eff}$ plane. We removed stars with rotation periods significantly longer than the Pleiades envelope at a given effective temperature since such stars are likely older. We then constructed a color–magnitude diagram (CMD) and used the \texttt{Glue} software \citep{2015ASPC..495..101B, robitaille_2019_3385920} to manually remove stars lying below the single-star locus of the Pleiades. A position below the cluster sequence typically indicates a fainter, and thus older, star at the same color, inconsistent with Pleiades membership. We additionally removed any stars that are overluminous in the CMD. These stars are likely binaries, which will influence both the radial velocity and rotation period measurements. Applying both the rotation and CMD cuts reduced our sample to 1,075 high-confidence members. These filters help eliminate field interlopers and ensure that the final radial velocity-only sample reflects a population consistent with the cluster’s age. Only 179 of the initial 1,254 stars had rotation periods or positions on the CMD that were removed (inconsistent with the Pleiades). 

The effects of these cuts are shown step-by-step in Figure~\ref{fig:quality_cuts}. Notably, the extended spatial structure of the complex is apparent even before applying HDBSCAN and remains visible after filtering on rotation and photometry, underscoring the coherence of the Pleiades' extended population. 


\begin{figure*}
    \centering
    \includegraphics[width=1\linewidth]{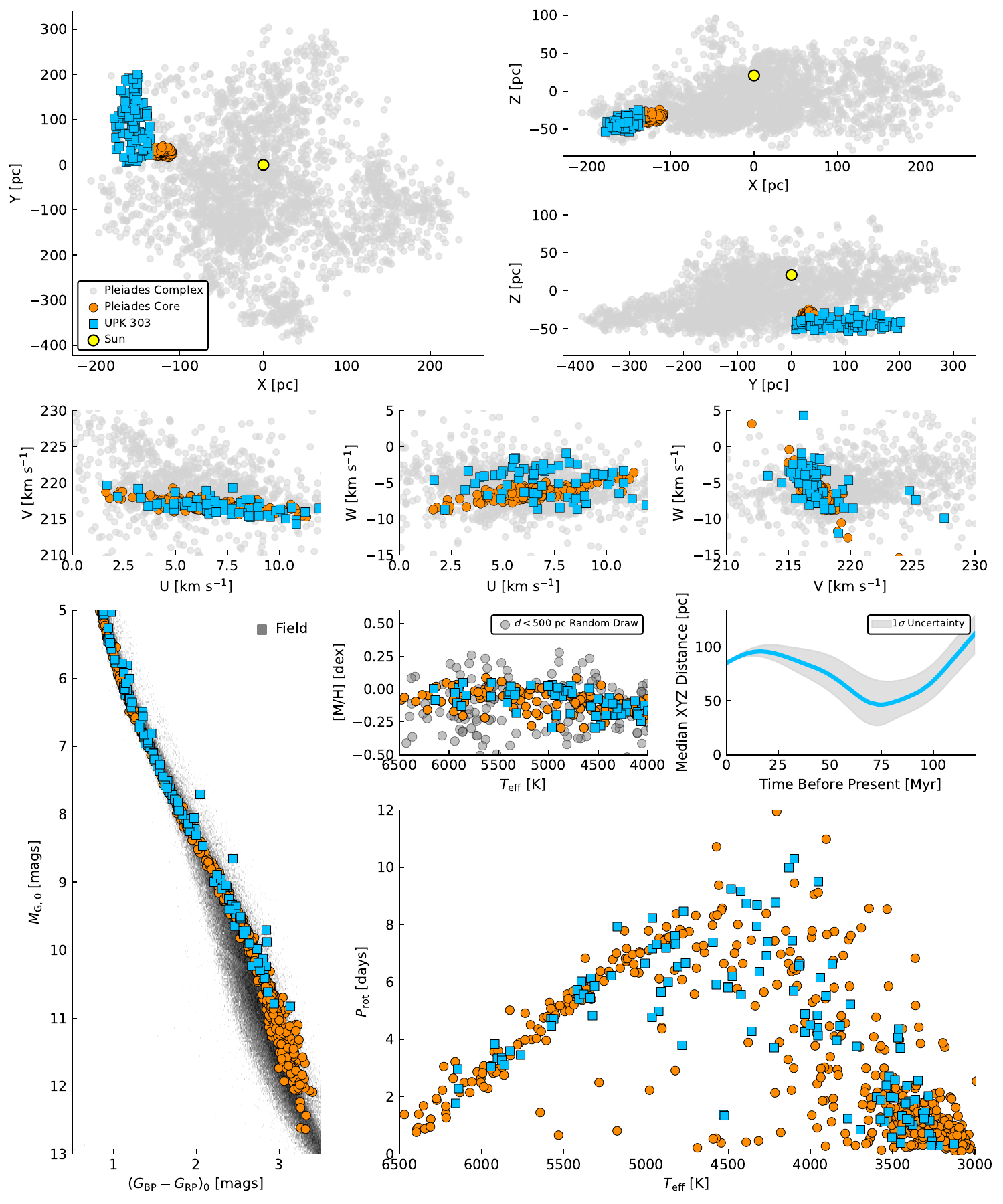}
    \caption{\textbf{UPK 303 and surrounding stars compared with the Pleiades}. Galactocentric positions (XY, XZ, YZ; top row) show that UPK 303 lies along a continuous spatial sequence extending from the Pleiades core. The second row shows Galactocentric velocities (UVW). The color–magnitude diagram (bottom left) and the rotation period–$T_{\mathrm{eff}}$ plane (bottom right) demonstrate that both structures lie along the same stellar sequence, consistent with a common age. Metallicity spread (bottom middle) show that UPK 303 members differ from a random sample of nearby field stars and are comparable to stars in the Pleiades core. The bottom-right panel shows the past separation between UPK 303 and the Pleiades, computed via 5000 Monte Carlo orbit integrations. Together, the spatial, kinematic, photometric, rotational, chemical, and dynamical similarities support a physical association between UPK 303 and the Pleiades.}
    \label{fig:upk_303}
\end{figure*}

\begin{figure*}
    \centering
    \includegraphics[width=1\linewidth]{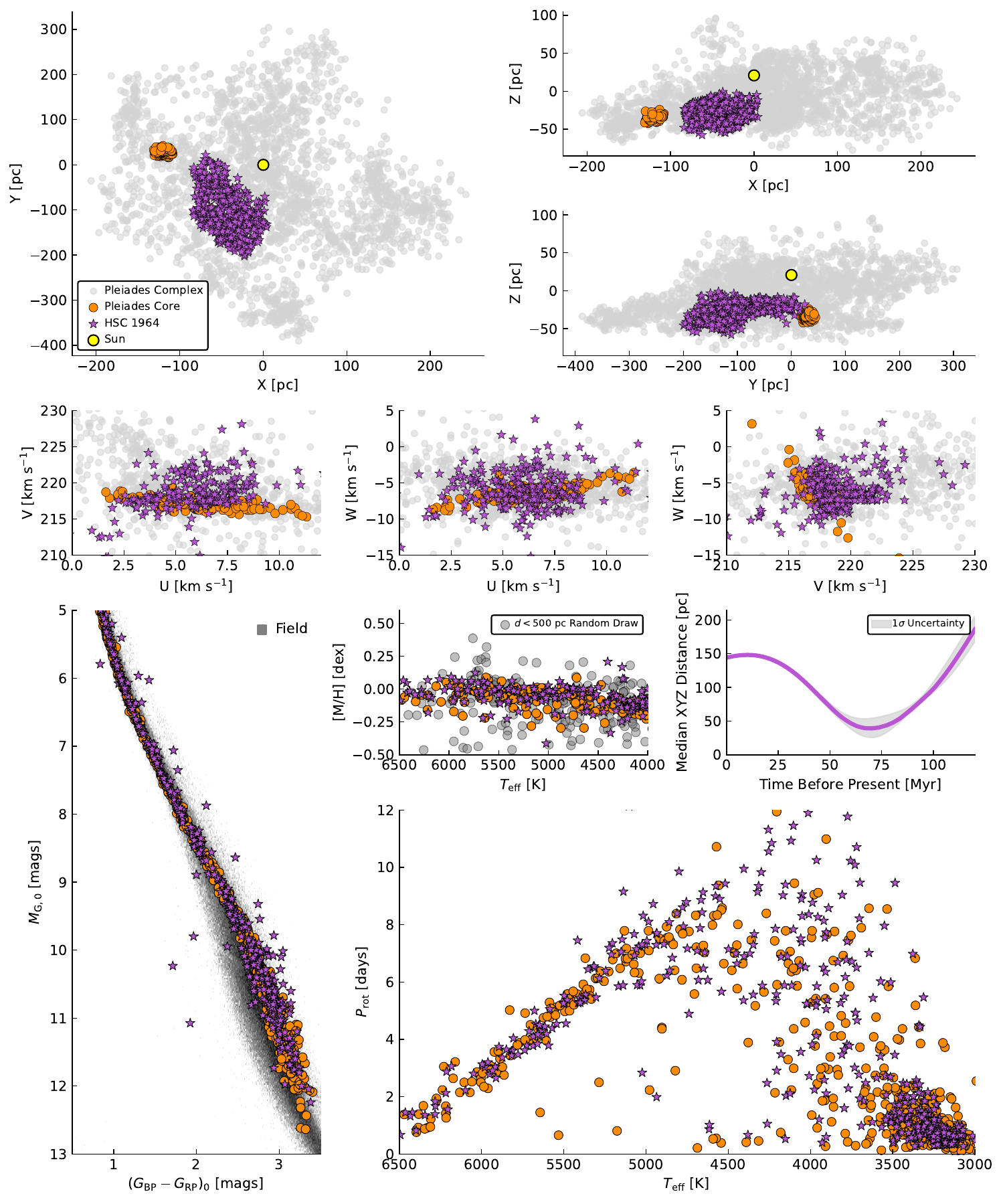}
    \caption{\textbf{HSC 1964 and surrounding stars compared with the Pleiades.} The structure of the plot is the same as in Figure \ref{fig:upk_303}. Like UPK 303, kinematics, age diagnostics, metallicity, and back-integrations all provide independent lines of evidence that HSC 1964 and the core of the Pleiades share a common origin.}
    \label{fig:hsc_1964}
\end{figure*}

\begin{figure*}
    \centering
    \includegraphics[width=1\linewidth]{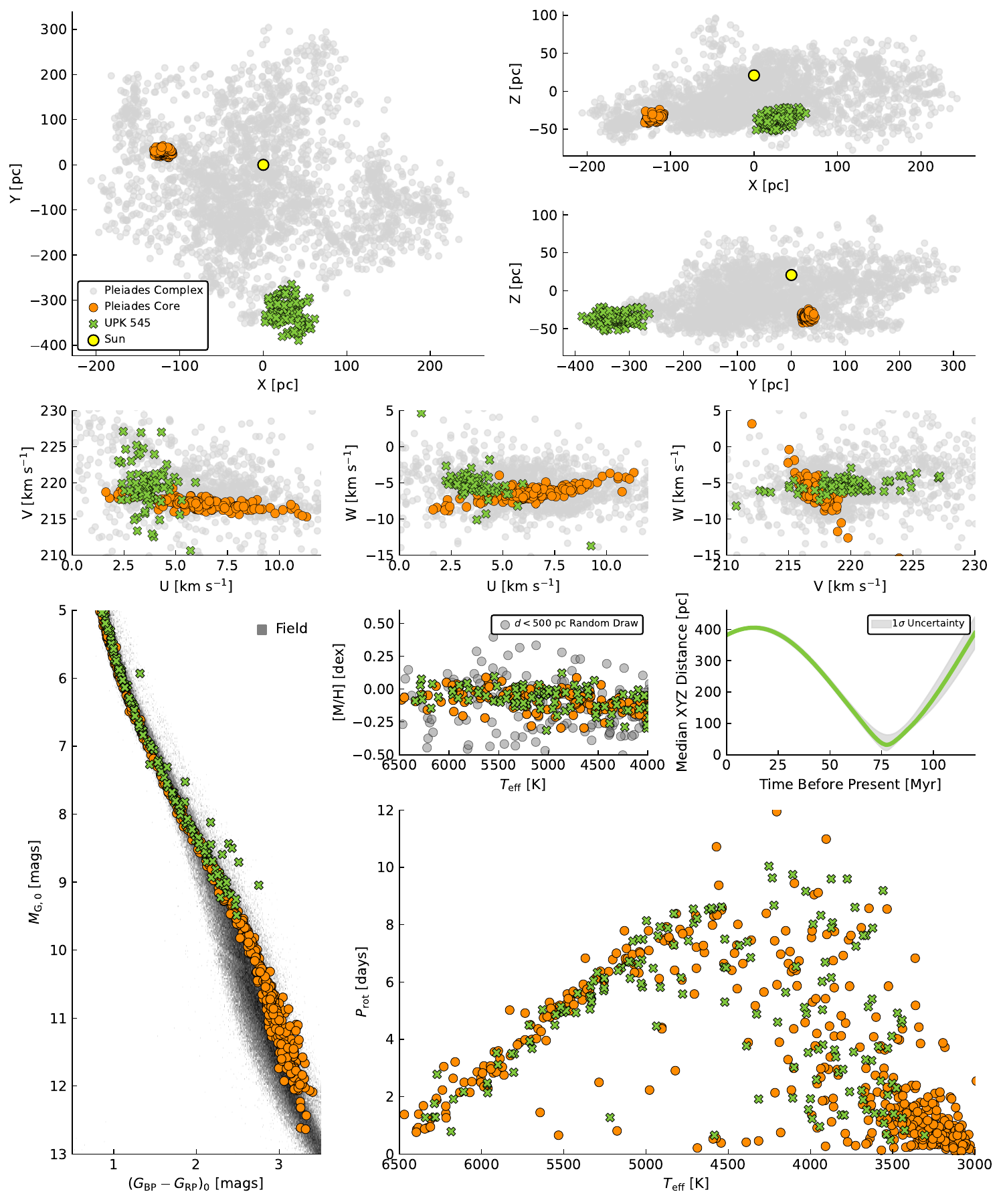}
    \caption{\textbf{UPK 545 and surrounding stars compared with the Pleiades.} The structure of the plot is the same as in Figure \ref{fig:upk_303}.  UPK 545 is further from the Sun than the other groups, leading to a dearth of M dwarfs in the CMD and \prot-\teff{} diagram, as well as a larger spread in velocities (larger velocity uncertainties), see Section \ref{subsec:upk_545} for more details. All available information again points towards a common origin with the Pleiades. }
    \label{fig:upk_545}
\end{figure*}

\begin{figure*}
    \centering
    \includegraphics[width=1\linewidth]{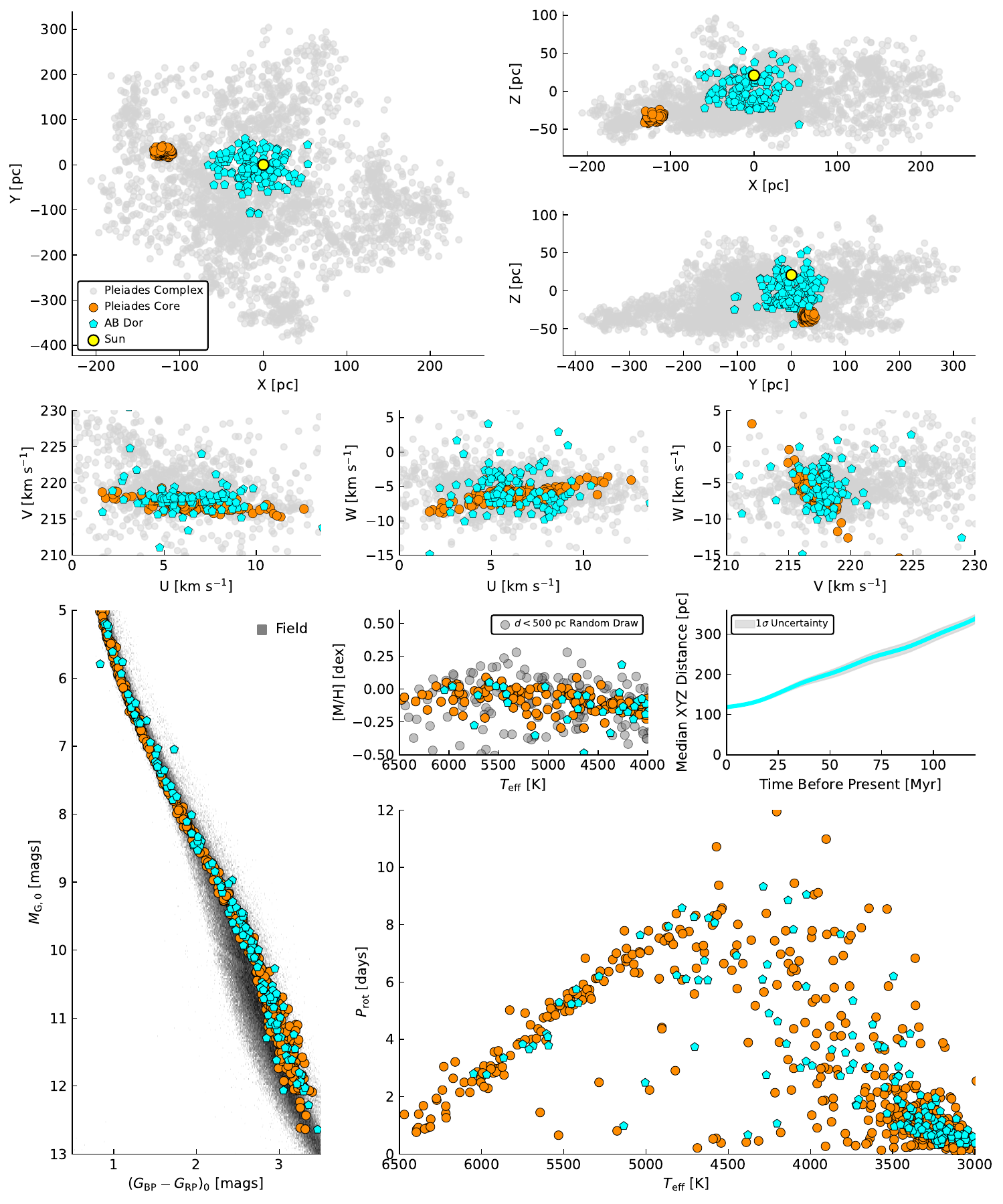}
    \caption{\textbf{The AB Doradus moving group compared with the Pleiades.} The structure of the plots is the same as in Figure~\ref{fig:upk_303}. Similar to other regions, kinematics, metallicity, and the CMD show evidence that AB Dor is related to the Pleiades core. AB Dor's back integration is not as definitive as for UPK~303, HSC~1964, and UPK~545. }
    \label{fig:abdor}
\end{figure*}

\begin{figure*}
    \centering
    \includegraphics[width=1\linewidth]{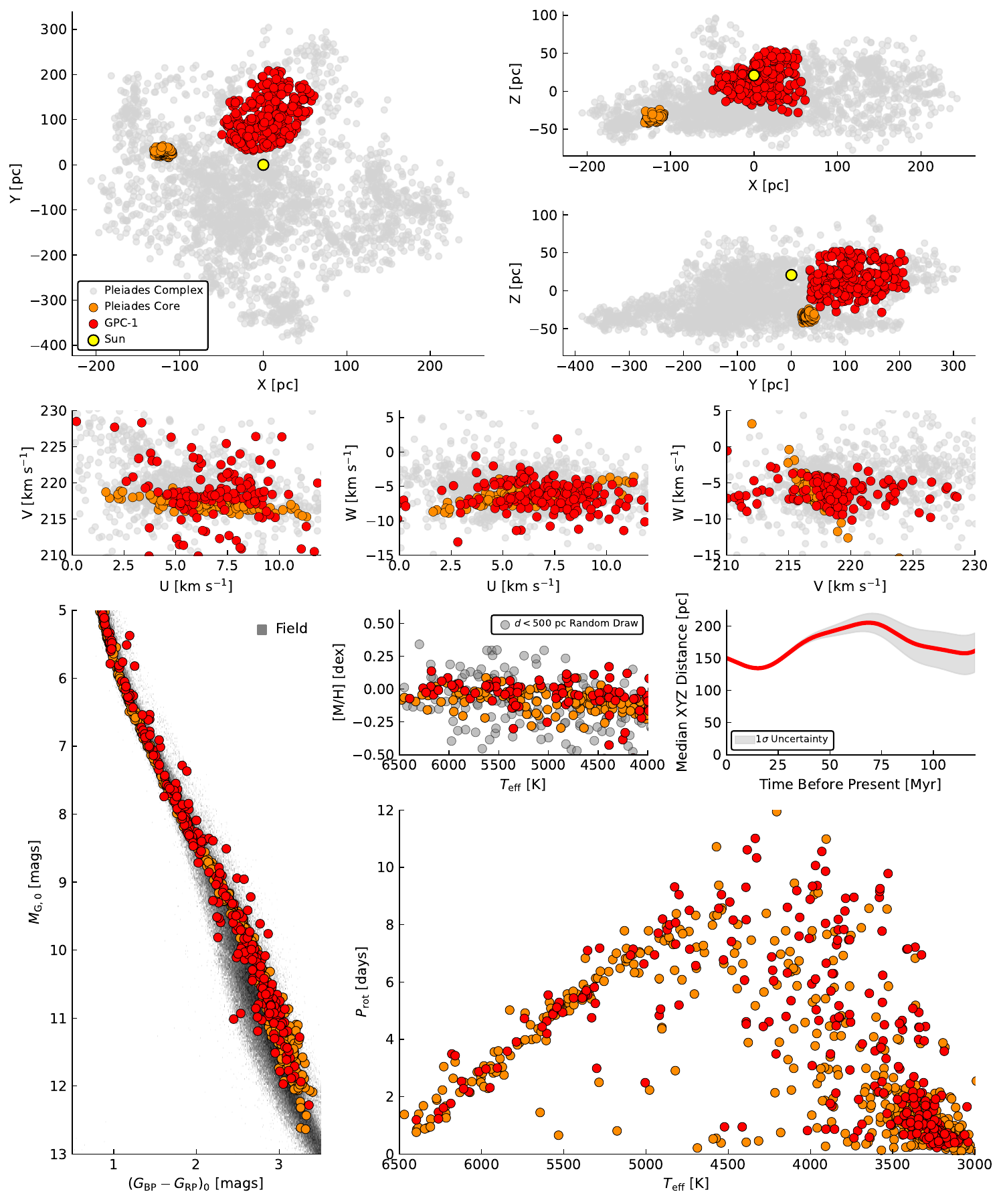}
    \caption{\textbf{The diffuse \upperright{} region compared with the Pleiades.} The structure of the plot is the same as in Figure \ref{fig:upk_303}. \upperright{} appears to have a large population of stars that are coeval with the Pleiades, though there is more scatter than in the CMD and rotation sequences than for UPK 303, HSC 1964, and UPK 545, and the back integration suggests that this region and the Pleiades were not closer in the past. This suggests a population that could be associated with the Pleiades, but that could also be mixed with the field or other young populations. }
    \label{fig:upper_right}
\end{figure*}

\begin{figure*}
    \centering
    \includegraphics[width=1\linewidth]{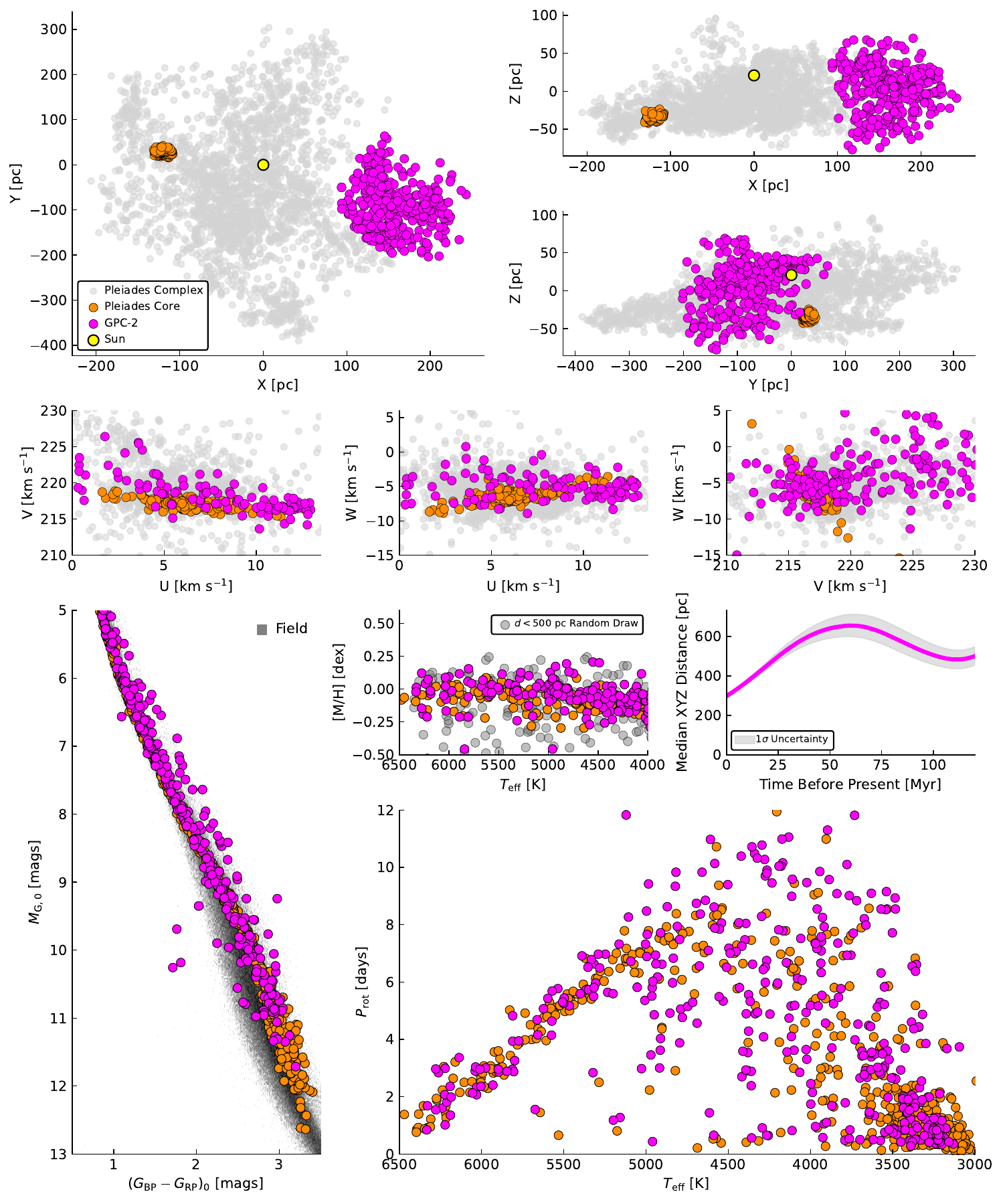}
    \caption{\textbf{The diffuse \lowerright{} region compared with the Pleiades.} The structure of the plot is the same as in Figure \ref{fig:upk_303}. Similar to the \upperright\ region, this region displays a large population of stars coeval with the Pleiades, albeit with a large amount of scatter in the CMD and rotation diagrams. The back integration is similarly ambiguous. As with \upperright{}, we suspect this region contains a combination of stars associated with the Pleiades, field stars, and other young populations.}
    \label{fig:lower_right}
\end{figure*}

\subsection{Expanding the Membership List} \label{subsec:comove}

The initial target list was restricted to stars with \gaia\ DR3 radial velocities. To expand the sample of candidate members beyond those with full six-dimensional kinematic information, we employed a customized version of the \texttt{Comove} algorithm\footnote{\url{https://github.com/adamkraus/Comove}}. Full details of the algorithm are provided in \citet{tofflemireTESSHuntYoung2021}, but to briefly summarize, \texttt{Comove} evaluates whether a star is kinematically consistent with a fiducial source by comparing its observed tangential velocity ($v_{\rm{tan,obs}}$) to the velocity expected if it were comoving with that source ($v_{\rm{tan,predicted}}$). \texttt{Comove} then selects stars based on user-provided thresholds in the tangential velocity offset ($v_{\rm{tan,off}} = |v_{\rm{tan,obs}}-v_{\rm{tan,predicted}}|$) and physical separation. 

As can be seen in Figures~\ref{fig:quality_cuts} and \ref{fig:groups}, the stellar density varies over the complex. The same effect is visible in kinematic space --- some groups show greater velocity dispersions. Thus, a single cutoff in $v_{\rm{tan,off}}$ and distance would over-select field stars in some regions and under-select members in others. To account for this, we set the threshold for each of the 1,075 targets using its fifth nearest neighbor within the sample of 1,075 high-quality members (Section~\ref{subsec:hdbscan}). This method automatically selects cutoffs appropriate for the density of the region. Other clustering studies \citep{kerrStarsPhotometricallyYoung2021, 2023ApJ...954..134K} have shown that similar methods successfully expand membership lists. We set a maximum cutoff of 3\,\kms\ in $v_{\rm{tan,off}}$ to avoid over-selection of non-members near the group edges. After combining all \texttt{Comove} runs, we retained a sample of 10,080 stars. 





The 10,080 star sample shows evidence of over-selection in the form of excess spread in the CMDs and rotation period sequence. Also, the next part of our analysis requires homogeneous assignment of membership probabilities (see Section~\ref{sec:bayes}), which is difficult given the variations in methods used to select members with and without radial velocities. As such, we did one final run through HDBSCAN. We used the same HDBSCAN setup as before, except this time we do the clustering on the three spatial dimensions (X, Y, Z) and $v_{\rm{tan,off}}$ (from \texttt{Comove}). After this final round of clustering, we were left with 6,961 candidate cluster members.

\section{Bayesian Membership Calculation} \label{sec:bayes}

A common limitation of clustering-based searches for stellar associations is their reliance on spatial and kinematic parameters---typically some combination of sky coordinates, distance, sky motions, and radial velocity---while omitting direct age constraints. Studies such as \citet{2023ApJ...954..134K} pre-filtered stars based on photometric indicators of youth (e.g., selecting those younger than 50 Myr) prior to clustering, but do not directly include age in their clustering or membership probabilities. In the absence of direct age constraints during the clustering phase, resulting associations generally require additional validation to confirm youth, typically through rotation, CMD position, or lithium diagnostics \citep[e.g.,][]{boumaRotationLithiumConfirmation2021, Boyle2023, Thao2024a}. This makes assigning membership probabilities for each star more challenging, especially absent a massive follow-up effort. 

Here, we explicitly incorporate stellar age into the membership assessment by incorporating rotation periods into a Bayesian framework. The goal is to assign membership probabilities based on both kinematics and age (estimated from the star's rotation). We derive the equations necessary for this membership calculation and present how each parameter used in the equations is calculated in Appendix~\ref{sec:full_bayes}.

We applied our derived equations and their  parameters as calculated in Appendix~\ref{sec:full_bayes} to our list of 6,961 candidate cluster members. This returns a membership probability bounded between 0 and 1 for every star in the list. Given that our calculation takes into account both the star's rotation period and its kinematic similarity with the Pleiades, we are able to calculate membership probabilities even in the absence of a measured rotation period, as might happen with a star with a pole-on orientation.
We adopted a final selection threshold of \mbox{$P(M \mid P_{\mathrm{rot}}, T_{\mathrm{eff}}) > 0.5$}, which yielded 3,091 stars with rotation-based support for membership.

\section{Results} \label{sec:results}

Our clustering and membership calculations reveal a set of coeval stars spanning nearly 600 pc. The full association contains 3,091 stars with $T < 16$. This includes at least seven known groups, including the Pleiades, UPK 303 \citep{2019JKAS...52..145S}, Theia 301\footnote{\citet{Hunt2023} lists Theia 301 as two groups; HSC 1580 and HSC 1964.} \citep{kounkelUntanglingGalaxyII2020}, Theia 163/UPK 545, OSCN 99 \citep{2023ApJS..265...12Q}, and the AB Dor moving group. The two remaining groups have no known literature counterparts; we refer to these as GPC-1 and GPC-2 (Figures~\ref{fig:upper_right} and~\ref{fig:lower_right}).  A top-down view of our membership list and overlapping groups is shown in Figure \ref{fig:groups}.

We split the \complex{} into sub-regions based on the literature association whose spatial and kinematic signature overlaps it most strongly. For a given region, we then include (1) all stars in that cluster that are also recovered in our membership list and (2) any additional stars that lie within the same three-dimensional volume defined by the cluster’s extent. We defined the GPC-1 and GPC-2 regions (Figure~\ref{fig:groups}) using manual spatial boundaries in order to select candidate members and assess their properties.

Basic data for each of the six groups are listed in Table~\ref{tab:summary_table} and described in the following section, where we also discuss each of the six groups in detail.

\subsection{Validating the \complex} \label{subsec:validation}

Unsupervised clustering methods like \added{HDBSCAN} run the risk of grouping stars with similar velocities that do not share a common origin. In addition to random effects, astrophysical overdensities can arise from non-symmetric perturbations in the Milky Way's potential \citep{Grand2015} and from dwarf galaxy mergers \citep{Helmi2020}. Known examples of random field stars with similar kinematics include the classical Pleiades and Hyades groups defined by UV-based clustering studies \citep{Eggen1975,Famaey2008} and some of the older Theia groups \citep{Zucker2022b,Barber2023}. 
To address this challenge, we perform a suite of tests to assess co-evality, chemical abundances, and a common traceback (origin) for the main components of the complex as well as its individual members. The tests were as follows.


\textbf{Color-magnitude position:} A collection of single-age, single-metallicity stars should create a color-magnitude sequence much tighter than the field. The remaining complications are binarity, uncorrected extinction variations (Section~\ref{subsec:reddening}), and photometric errors. The latter two will be a bigger problem for more distant groups, and all groups will have some non-member interlopers (see Section~\ref{subsec:contamination}). Otherwise, these issues will have a comparable impact on the group as the Pleiades core. Hence we compare each group CMD to the canonical Pleiades members and to a set of field stars from the \gaia\ Catalog of Nearby Stars \citep{2021A&A...649A...6G}. We note that although a CMD cut was applied to select the core of each group (Section~\ref{subsec:hdbscan}), no such cut was applied when using \texttt{Comove} (Section~\ref{subsec:comove}). Since \texttt{Comove} is responsible for most of the member stars, this remains a strong test. For 100 Myr stars, the CMD should be most elevated above the field sequence for $2 < G_{\mathrm{BP}} - G_{\rm RP}< 3$. We compute the average absolute \gaia\ $M_{\rm G}$ in 0.1 magnitude wide color bins over this range and report the average number of standard deviations in this range that each group's sequence lies above the field sequence.

\textbf{Rotation sequence:} As with the CMD, a co-eval young population should have a \prot-\teff{} sequence set by standard rotational evolution \citep[e.g.,][]{Bouma2023}. Our Bayesian membership probability biases the rotation sequence of selected members towards Pleiades, so this test is less diagnostic than the others. However, we provide the \prot-\teff{} comparison to the Pleiades for each sub-group for reference.

\textbf{Excess Variability Age (EVA):} \citet{Barber2023} present a test to measure the age of a cluster based on the excess noise in \gaia\ photometry. The assumption is that noise above measurement precision is driven mostly by photometric variability combined with a relation between age and spot coverage \citep{Morris2020}. Ages derived this way are subject to large uncertainties, but the metric is only weakly impacted by field interlopers and is separate from the CMD and \teff-\prot{} tests. 

\textbf{Chemical Abundances:} Members of individual open clusters are expected to share abundance patterns across many elements because most nucleosynthetic products are well‐mixed in the progenitor molecular cloud (e.g., \citealt{2010ApJ...713..166B, 2024MNRAS.529.2483B}). Chemical consistency can provide support for a common origin and a common age \citep{Miller2025}

Although no single spectroscopic survey uniformly covers the full extent of the Greater Pleiades Complex, we adopt the homogenized bulk metallicities ([M/H]) from \citet{2023ApJS..267....8A}, who derived [M/H] from \gaia\ BPRP spectra using APOGEE-trained XGBoost models. The \citet{2023ApJS..267....8A} metallicity scale shows a 'bend' at \teff$<$4500\,K, such that [M/H] is systematically underestimated for cooler stars. Given this known bias, we compare the [M/H]-\teff{} sequence of each sub-group to that of the Pleiades core and a control sample of stars randomly drawn from the parent population. 

\begin{figure*}[!t] 
  \centering
  \includegraphics[width=0.95\textwidth]{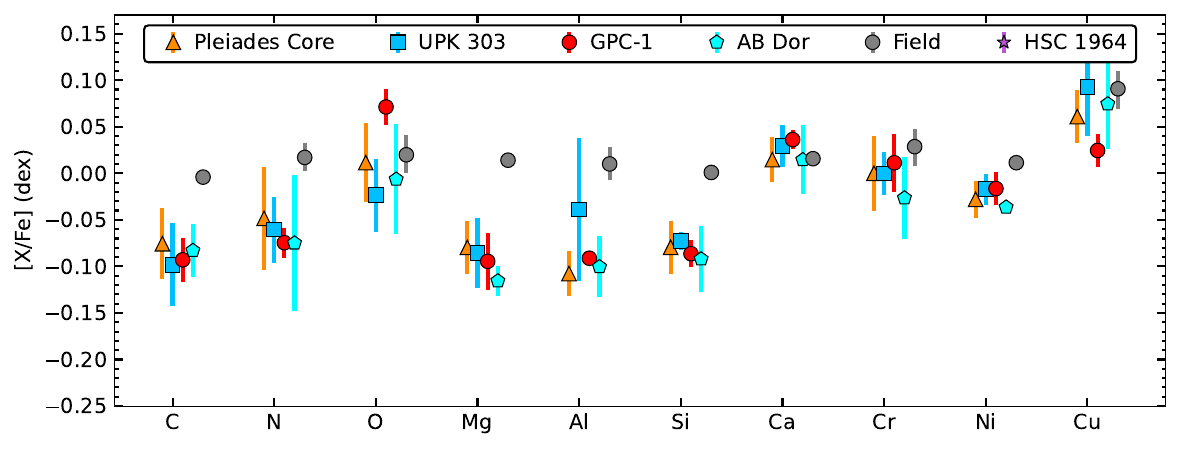}
  \includegraphics[width=0.95\textwidth]{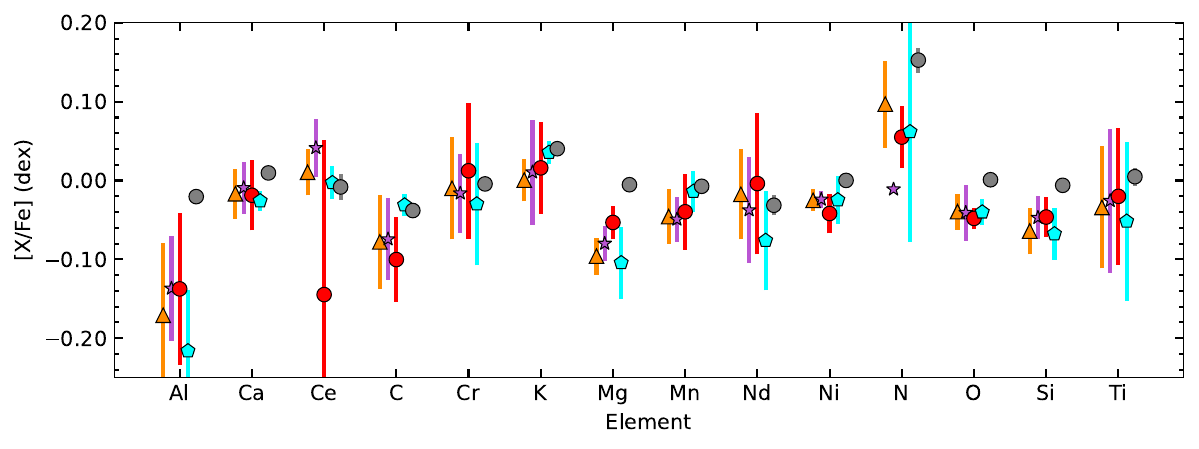}
  \caption{\textbf{Individual elemental abundances support a common formation between the Pleiades core, UPK 303, HSC 1964, and the diffuse \upperright{} region}. \textit{Top:} Median $[\mathrm{X}/\mathrm{Fe}]$ and 1$\sigma$ dispersion for various elements, as measured from LAMOST DR7 medium-resolution spectra for the Pleiades core, UPK 303, and the \upperright{} region. Gray points show a random field sample drawn from stars within 500 pc and $T<16$.
  \textit{Bottom:} Same as above but using \added{Milky Way Mapper (SDSS DR19) high-resolution abundances for the Pleiades core, HSC 1964, and the \upperright{} region. We exclude P, V, and Cu because measurements of these elements were marked as having poor quality in \cite{2025AJ....170...96M}. We additionally only include stars with $3500 < T_{\rm eff} < 6000$ to match the parameter ranges in \cite{2025AJ....170...96M} where measurements are optimal.}
  In both panels, the close agreement of our substructures with the Pleiades core across multiple elements --- and their divergence from the broader field distribution (especially in C, Al, Mg, and Si) --- suggests a common origin. Note that UPK 545, HSC 1964, and the \lowerright{} region lack coverage in LAMOST DR7 (top), and UPK 303, UPK 545, and the \lowerright{} region are absent from \added{MWM DR19} (bottom).}
  \label{fig:abundances}
\end{figure*}

LAMOST DR7 \citep{2012RAA....12.1197C} and \added{Sloan Digital Sky Survey DR19 Milky Way Mapper (MWM) data \citep{1973ApOpt..12.1430B, 2006AJ....131.2332G, 2013AJ....146...32S, 2019PASP..131e5001W, 2025arXiv250706989K}} also provide a suite of elemental abundances for a subset of stars in each group. Figure~\ref{fig:abundances} compares median $[\mathrm{X}/\mathrm{Fe}]$ measurements for the Pleiades core and four of the six substructures from these surveys. The two regions not shown (UPK~545 and \lowerright{}) do not have sufficient coverage in either survey. 

\textbf{Kinematic back integration:} If the groups formed together, they should have been much closer in the past. We first corrected our \gaia\ DR3 radial velocities for magnitude-dependent systematic effects as described in \citet{2023A&A...674A...5K} and removed stars with \gaia\ RUWE $>1.2$ since binarity can influence \gaia\ radial velocity measurements. We then back-integrated the median present-day phase space positions and velocities of stars in each group relative to the median positions and velocities of the Pleiades core using \texttt{gala}\footnote{\url{https://github.com/adrn/gala}} \citep{gala, adrian_price_whelan_2024_13377376} and the \texttt{MilkyWayPotential2022} Milky Way potential. We ran this out to the age of the Pleiades (125\,Myr), calculating the 3D separation between the Pleiades core and each group at each timestamp. For uncertainties, we ran 5000 Monte Carlo trials, sampling from uncertainties in the astrometry and radial velocities, before converting to Galactocentric coordinates and taking the median of each dimension in Galactocentric coordinates.

\added{Although giant molecular clouds can span several hundred parsecs \citep{2011ApJ...729..133M}, allowing for the possibility that stars formed at large initial separations from the Pleiades core, a closer 3D separation in the past still serves as supporting evidence for a shared origin.} 



\begin{deluxetable*}{lcccccccccccc}
\tablecaption{Properties of Groups in the Greater Pleiades Complex}
\tabletypesize{\footnotesize}
\tablehead{
\colhead{Group Name} & 
\colhead{[M/H] Spread} & 
\colhead{\# Elements} & 
\colhead{Traceback} & 
\colhead{Traceback} & 
\colhead{EVA Age} & 
\colhead{U} & 
\colhead{V} & 
\colhead{W} & 
\colhead{X} & 
\colhead{Y} & 
\colhead{Z} \\
\colhead{} & 
\colhead{dex} & 
\colhead{matched$^a$} & 
\colhead{pc} & 
\colhead{Myr$^b$} & 
\colhead{Myr} & 
\colhead{\kms} & 
\colhead{\kms} & 
\colhead{\kms} & 
\colhead{pc} & 
\colhead{pc} & 
\colhead{pc}
}
\startdata
Pleiades Core  & 0.15 & ... & ... & ... & $88^{+19}_{-14}$ & $6.10^{+1.66}_{-1.39}$ & $217.22^{+0.65}_{-0.71}$ & $-6.38^{+0.83}_{-0.79}$ & $-120.37^{+3.48}_{-3.60}$ & $29.33^{+3.95}_{-3.98}$ & $-33.12^{+2.67}_{-2.88}$ \\
UPK 303 & 0.20 & 10/10 & $43^{+21}_{-19}$ & $73^{+6}_{-4}$ & $128^{+43}_{-28}$ & $6.53^{+3.26}_{-8.78}$ & $216.83^{+2.47}_{-1.22}$ & $-4.21^{+1.58}_{-3.34}$ & $-156.10^{+5.84}_{-12.46}$ & $110.72^{+35.73}_{-52.63}$ & $-43.89^{+6.21}_{-3.60}$ \\
AB Dor & 0.16 & 15/15 & ... & ... & $39^{+32}_{-12}$ & $6.70^{+1.93}_{-11.45}$ & $217.43^{+0.79}_{-1.67}$ & $-6.80^{+3.40}_{-1.98}$ & $-25.37^{+13.35}_{-21.36}$ & $16.53^{+19.76}_{-23.33}$ & $2.39^{+13.40}_{-10.45}$ \\
HSC 1964/Theia 301 & 0.14 & 14/14 & $39^{+12}_{-12}$ & $69^{+2}_{-2}$ & $73^{+14}_{-21}$ & $6.41^{+1.98}_{-2.56}$ & $218.14^{+6.02}_{-1.34}$ & $-5.79^{+3.36}_{-2.87}$ & $-33.85^{+22.52}_{-16.07}$ & $-106.80^{+42.69}_{-34.92}$ & $-20.83^{+7.54}_{-6.93}$ \\
UPK 545/Theia 163 & 0.14 & ... & $31^{+30}_{-18}$ & $78^{+1}_{-2}$ & $82^{+26}_{-16}$ & $3.52^{+0.73}_{-0.52}$ & $219.01^{+2.14}_{-3.59}$ & $-5.25^{+0.92}_{-0.75}$ & $28.53^{+6.03}_{-12.74}$ & $-320.95^{+10.60}_{-13.38}$ & $-35.92^{+4.18}_{-3.32}$ \\
GPC-1         & 0.16 & 15/15 & $134^{+2}_{-11}$ & $15^{+11}_{-9}$ & $132^{+38}_{-25}$ & $7.45^{+6.06}_{-2.60}$ & $219.29^{+24.05}_{-3.68}$ & $-6.08^{+2.10}_{-2.17}$ & $10.15^{+29.39}_{-28.89}$ & $118.14^{+40.42}_{-33.18}$ & $13.20^{+20.41}_{-15.42}$ \\
GPC-2         & 0.19 & ... & $483^{+47}_{-43}$ & $115^{+5}_{-3}$ & $66^{+20}_{-13}$ & $0.94^{+10.94}_{-18.47}$ & $223.43^{+11.50}_{-6.98}$ & $-4.32^{+4.77}_{-2.53}$ & $148.94^{+42.79}_{-27.53}$ & $-101.76^{+71.22}_{-65.07}$ & $-2.32^{+31.67}_{-30.93}$ \\
Field      & 0.38 & 11/15 & ... & ... & ... & ... & ... & ... & ... & ... & ... \\
\enddata
\footnotesize{$^a$ Some groups do not have coverage in every element; we only consider elements with at least two non-null measurements for that group.}\\
\footnotesize{$^b$ The time of closest approach is given as the time before present day.}\\
\added{\footnotesize{\textbf{Column Descriptions:} [M/H] Spread gives the dispersion in bulk metallicity within each group. \# Elements shows how many elemental abundances matched the Pleiades core within 2$\sigma$. Traceback (pc) is the closest 3D separation from the Pleiades core during back-integration, and Traceback (Myr) is the corresponding time of minimum separation. EVA Age is the age estimate from the Excess Variability Age method of \citet{Barber2023}. $U,V,W$ are Galactocentric velocity components (positive $U$ toward the Galactic center, $V$ in the direction of Galactic rotation, $W$ toward the north Galactic pole). $X,Y,Z$ are Galactocentric Cartesian positions (positive $X$ toward the Galactic center, $Y$ in the direction of Galactic rotation, $Z$ toward the north Galactic pole).}}\\
\label{tab:summary_table}
\end{deluxetable*}

\subsubsection{UPK 303} \label{subsec:upk_303}

UPK 303 was proposed as a potential tidal tail of the Pleiades by \citet{2025A&A...694A.258R}. We independently recover this structure and present our comparative analysis in Figure~\ref{fig:upk_303}. In Galactocentric spatial coordinates (XYZ), UPK 303 appears continuous with the Pleiades core. The Galactocentric velocities (UVW) show significant overlap, with only some offset in $W$. \citet{kounkelUntanglingGalaxyLocal2019} lists three stars as part of Theia 369 that overlap with \citet{Hunt2023}'s UPK 303 membership list and that are recovered in our membership list.

Members of UPK 303 and the Pleiades core lie along the same single-star sequence in both the color–magnitude diagram (CMD) and the \prot–$T_{\rm eff}$ plane and we can rule out these stars having a field CMD sequence at 4.2$\sigma$. The inferred [M/H] distribution of UPK 303 closely matches that of the Pleiades core, and is distinct from a random draw of field stars at similar distances. Detailed abundances also match for all 10 elements tested (Figure~\ref{fig:abundances}). 

Our back integration suggests the median closest approach between UPK 303 and the Pleiades core occurred $73^{+8}_{-4}$~Myr ago at a separation of $43^{+21}_{-19}$~pc.

Taken together---spatial and kinematic continuity, overlapping age diagnostics, consistent metallicities, and historical proximity---these results strongly suggest that UPK 303 and the Pleiades share a common origin.

\subsubsection{HSC 1964 (Theia 301)}\label{subsec:hsc_1964}

The stars in the vicinity of HSC 1964 form a large, diffuse stellar aggregate located approximately 160 pc from the core of the Pleiades. This structure was also identified as part of Theia 301 by \citet{kounkelUntanglingGalaxyLocal2019}, with 85 stars overlapping between the two membership lists that are also recovered in our membership list. We also note that eight stars from our membership list that \citet{kounkelUntanglingGalaxyLocal2019} list as part of Theia 301 are recovered in HSC 1580 by \citet{Hunt2023}. 

In spatial coordinates, the structure is elongated and detached from the core, but lies along a continuous stellar bridge (Figure \ref{fig:hsc_1964}, top panels). In velocity space, the HSC 1964 population exhibits UVW velocities consistent with those of the core (Figure \ref{fig:hsc_1964}, middle panels).

The stellar population of HSC 1964 has CMD and \teff-\prot{} sequences that are indistinguishable from the core and we can rule out these stars have a field CMD sequence at 2.2$\sigma$. The metallicity sequence also matches that of the core with far less spread than the field draw. 

Our backward orbital integration shows that HSC 1964 approached to within $37^{+12}_{-12}$ pc of the Pleiades core roughly $69^{+2}_{-2}$ Myr ago. This minimum separation is significantly smaller than its current distance, and is much less than the separation between subgroups in the Orion complex (see Figures~\ref{fig:fig1} and Figure~\ref{fig:orion_back_int}). 

Taken together, the kinematics, stellar ages, chemical abundances, and orbital history of HSC 1964 point to a shared formation history with the Pleiades. 

\subsubsection{UPK 545 (Theia 163)}\label{subsec:upk_545}

UPK 545 and surrounding stars form a diffuse stellar population located nearly 400\,pc from the core of the Pleiades. As with UPK 303 and HSC 1964, we identify a continuous chain of coeval, comoving stars that spatially links UPK 545 to the Pleiades core (see Figure~\ref{fig:upk_545}). Although it lies far outside the classical tidal radius, there is a bridge of stars to the core of the Pleiades via HSC 1964/Theia 301.

In velocity space, stars in UPK 545 exhibit UVW velocities that are slightly offset from the Pleiades' core. Individual members show more spread, but some of this is due to larger uncertainties (especially in $V$) due to a greater distance from the Sun. The CMD and the \prot-\teff{} diagram are both indistinguishable from the Pleiades. There is, unfortunately, a deficit of stars below 3500\,K in both diagrams, as such stars at this distance are $T>16$ and hence do not have usable \tess\ light curves. For the stars that do fall in the range of $2 < G_{\mathrm{BP}} - G_{\rm RP}< 3$, we can rule out a field-star CMD distribution at 2.0$\sigma$.

The metallicity distribution similarly matches that of the core, with [M/H] values incompatible with a randomly drawn field sample at the same distance. UPK 545 unfortunately did not have enough points in LAMOST or MWM for a comparison of detailed abundances. 

Orbital back integration reveals that UPK 545 passed within $31^{+30}_{-18}$ pc of the Pleiades core approximately $78^{+1}_{-2}$ Myr ago. This is highly constraining, as it is one of the most distant groups and yet has one of the closest approaches, placing it well within the spatial extent of the modern Pleiades. 

This close approach, combined with its present-day kinematic alignment and coeval stellar properties, argues strongly for a common formation history. Despite its current separation of nearly 400 pc, UPK 545 appears to be a dispersed remnant of the same star-forming event that produced the Pleiades.


\subsubsection{AB Doradus} \label{subsec:abdor}


The AB Doradus moving group \citep{2004ApJ...613L..65Z} has long been suggested to be related to the Pleiades \citep{Luhman2005}. The analysis by \citet{2007MNRAS.377..441O} in particular integrated the orbits of AB Dor members and the Pleiades backward in time, and found that the two groups reached a minimum separation roughly 120 Myr ago—consistent with most age estimates for the Pleiades. Since then, several studies have supported a link between AB Dor and the Pleiades \citep[e.g.,][]{2022AJ....163..289Z}.

To cross-match our results against previously reported AB Dor members, we retrieved candidate members from the Montreal Open Clusters and Associations (MOCA) database (J. Gagn'e et al., in preparation; \citealt{2024PASP..136f3001G}), which yielded 834 stars and brown dwarfs. 547 of these sources were detected in \gaia\ DR3 and met our criteria for consideration: $3000~\mathrm{K} < T_{\rm eff} < 6500~\mathrm{K}$ and $T < 16$. Among these 547 stars, 146 stars were recovered in our sample, of which 20 are marked as ``bona fide'' or ``highly-likely'' candidates members in the MOCA database. 

We performed our suite of comparative tests for AB Dor, and the results are shown in Figure~\ref{fig:abdor}. As observed in other regions, metallicities, color–magnitude diagram (CMD) positions, rotation periods, and kinematics all overlap with the Pleiades core, with the CMD sequence lying 4.8$\sigma$ above the field CMD sequence. Furthermore, detailed elemental abundances presented in Figure~\ref{fig:abundances} match with those of the Pleiades for all 15 elements studied.

However, results from orbit integration tests remain ambiguous. The results shown in Figure~\ref{fig:abdor} are nearly identical if we use all of the 146 candidates recovered in our sample or only the 20 ``bona fide'' or ``highly-likely'' members. To further test this scenario, we integrated orbits for each of our confirmed AB Dor members relative to the Pleiades core. Only one star yielded a minimum separation time exceeding 100 Myr in the past, indicating that our results do not reproduce the findings reported by \citet{2007MNRAS.377..441O}. We list the back integration results for AB Dor in Table~\ref{tab:summary_table} as null results since the back integration only shows AB Dor getting further away from the Pleiades core in the past. 

\subsubsection{\upperright}\label{subsec:upper_right}

In addition to the known regions analyzed above, we identified a previously unrecognized diffuse stellar population 150~pc away from the Pleiades core (Figure~\ref{fig:upper_right}). None of the clustering catalogs we cross-matched flagged this group as a distinct entity. For the purposes of this study, we refer to this structure as \upperright. 

Applying the same set of diagnostics used for UPK 303, HSC 1964, and UPK 545, we find that the \upperright{} members form a coherent spatial overdensity that, while not directly connected to the core, lies along the broader structural extension of the Pleiades complex. In velocity space, the population is clustered and overlaps with the Pleiades in UVW coordinates. The CMD shows a relatively tight sequence matching that of the Pleiades core and a sequence lying 2.8$\sigma$ above the field sequence. The estimated metallicities match the Pleiades, and all individual elements tested are $2\sigma$ consistent with Pleiades estimates.

The orbital history of this putative group is ambiguous. Our backward integration showed that the \upperright{} population has not been within $134^{+2}_{-11}$ pc of the Pleiades core over the past 120~Myr. While this separation is marginally smaller than its present-day distance ($\sim$150 pc), it does not represent the level of convergence seen in the other three regions. The broader uncertainties in the traceback also suggest a less definitive dynamical connection. A second minimum 115\,Myr ago is close to the commonly cited age of the Pleiades and could potentially become more prominent with a cleaner membership list in this region.

In summary, while the \upperright{} region presents some evidence for physical association with the Pleiades based on age, present-day kinematics, and chemistry, the kinematic traceback is less conclusive. Given the large spatial extent and the spread seen in $UVW$, this may be because the group is actually multiple sub-populations, or a real group with higher field contamination than the other populations.

\subsubsection{\lowerright}\label{subsec:lower_right}

We also identify a diffuse stellar population 300~pc from the Pleiades core (Figure~\ref{fig:lower_right}). This region partially overlaps with the open cluster OSCN 99, but spans a larger volume than the cluster itself. Because the broader structure has not been previously cataloged, we refer to it as \lowerright{}.

The stellar population within \lowerright{} appears broadly coeval with the Pleiades, though with greater scatter in both the CMD and the rotation--effective temperature diagram. The CMD sequence lies 3.9$\sigma$ above the field sequence. While a coeval sequence is still discernible, the broader spread suggests larger contamination in this region.

In UVW space, the \lowerright{} stars exhibit weaker kinematic coherence compared to the other four regions analyzed. While a cluster-like signature is visible in the VW projection, the overall distribution is more dispersed, and no strong overdensity is apparent in the U dimension. This stands in contrast to the tighter clustering observed for UPK 303, HSC 1964, UPK 545, and the \upperright{} region.

The stars in the \lowerright{} region show a metallicity distribution consistent with the Pleiades and inconsistent with that of the local field population, although, as with kinematics and CMD, the spread is larger than the Pleiades. 

Our backward integration indicates that the \lowerright{} region passed within $483^{+47}_{-43}$ pc of the Pleiades core approximately $115^{+5}_{-3}$ Myr ago. While this minimum occurs around the time the Pleiades was formed, the large spatial difference broad uncertainties limit the strength of any dynamical connection. Further cleaning of the membership list in that area could help clarify if this minimum represents a dynamical connection to the Pleiades.

OCSN 99 also appears in \citet{kounkelUntanglingGalaxyLocal2019} as Theia 368. If we back integrate just the stars that comprise OCSN 99, we find that it had a minimum separation of $\sim$200 pc from the Pleiades core 25 -- 50 Myr ago. Similar to the rest of this region, the evidence for a dynamical connection with the Pleiades is ambiguous.

In summary, the \lowerright{} region hosts a large population of stars with ages, metallicities, and kinematic properties similar to the Pleiades. However, the weaker kinematic coherence and ambiguous orbital history suggest that its connection to the Pleiades complex is less certain than for the other regions. It's possible that this region contains a number of stars that are related to the \complex{} (including OSCN 99) mixed with a large number of contaminants. Further follow-up, including improved membership constraints and spectroscopic age diagnostics, would help clarify the nature of this extended structure.

\subsection{Are Back Integrations Viable in This Situation?}

In Appendix \ref{sec:back_int_experiment}, we employ the methodology used in our analysis to integrate subgroups of the Orion Complex forward to show that \gaia\ kinematics are---under certain conditions---sufficient to trace resulting sub-populations back to their $\simeq$100\,Myr parent. We also show that this method gives reasonable ages for the epoch of closest approach, but systematically overestimates the separation. Here we discuss the independent issue of how likely it is to get back integration results like those we see by chance. 

To this end, we built a control sample, starting from our initial selection of stars moving within 5~km~s$^{-1}$ of the Pleiades core. We then excluded the 3,272 complex members recovered in our analysis, so that the sample contains mostly non-members. We also imposed additional quality cuts: radial velocity signal-to-noise $RV/\sigma_{RV} > 5$ and uncertainties $\sigma_{RV} < 1~\mathrm{km~s^{-1}}$. These cuts ensure more reliable orbit tracing, and help remove bona fide members we missed due to poor data quality. For each group, we then (1) compute its present-day separation from the Pleiades, (2) randomly sample a spatial location within 25~pc of that separation, (3) identify all stars within 10~pc of that location, and (4) back-integrate the median phase-space coordinates of the selected stars relative to the Pleiades. We repeat this process 100 times for each region, each time requiring that we sample a new region of space. We do not include GPC-1 and GPC-2 in this experiment because their back-integrations are not as well defined as those for UPK 303, HSC 1964, and UPK 545.

\begin{figure*}
    \centering
    \includegraphics[width=1\linewidth]{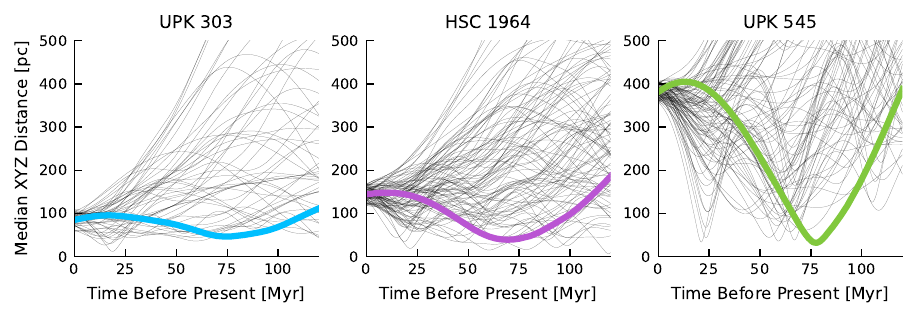}
    \caption{Back-integrated 3D separations between each group (UPK~303, HSC~1964, and UPK~545) and the Pleiades core as a function of time before the present. The colored, solid curve in each panel shows the same trajectory as in the diagnostic plots for each region above. Black curves show the same quantity for 100 randomly selected regions at similar present-day separations. The diagnostic value of our back integrations is supported by the fact that few random samples yield similarly close approaches, particularly at comparable times.}
    \label{fig:back_int}
\end{figure*}

Figure~\ref{fig:back_int} summarizes the outcome. We find that 14\%, 17\%, and 1\% of the random samples for UPK~303, HSC~1964, and UPK~545, respectively, exhibit a minimum separation from the Pleiades smaller than that of the actual group. These back integrations are accompanied by large uncertainties (see Appendix \ref{sec:back_int_experiment}), so a more accurate test is to look for regions that have a global minimum at close to the same time that the real cluster has a global minimum distance with Pleiades. If we only consider back integrations that have a minimum within $\pm10$ Myr of the time of minimum separation of each cluster, these numbers drop to 6\%, 4\%, and 2\%, respectively. 

These results indicate that the back integrations for these three groups are unlikely to occur by chance. Therefore, the observed close passages are not generic among stars with similar present-day kinematics and separations, and the back-integrated convergence provides meaningful evidence for a physical association with the Pleiades, especially when combined with age and abundance metrics. 

\subsection{Expansion Signatures}\label{subsec:expansion}

\begin{figure*}
    \centering
    \includegraphics[width=0.9\linewidth]{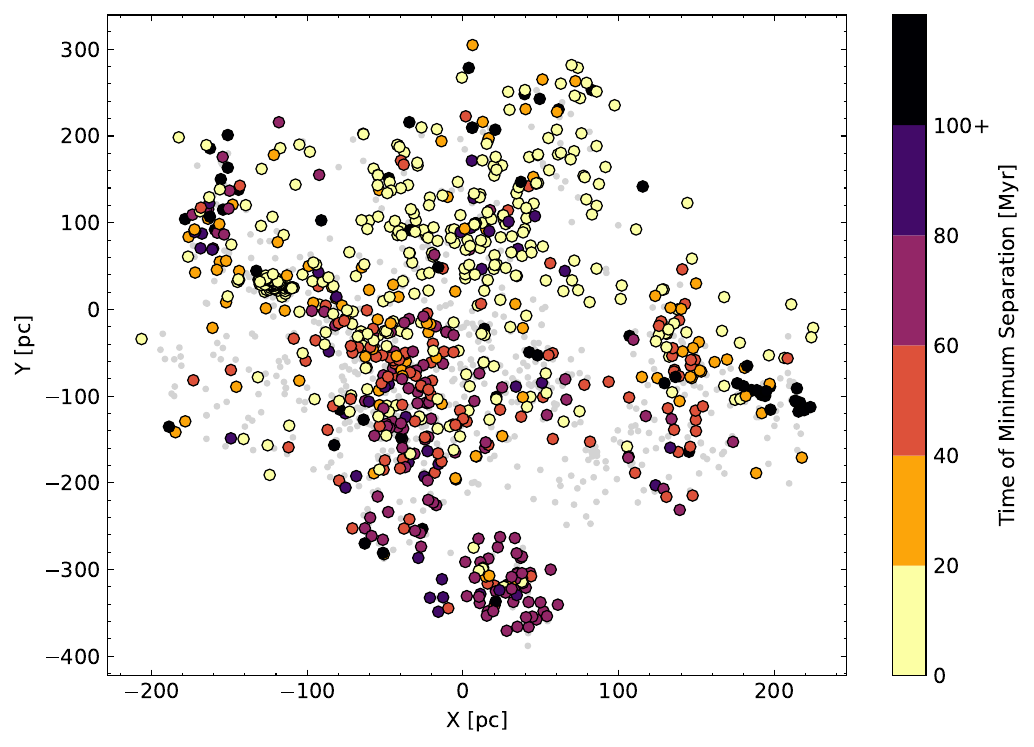}
    \caption{The time of minimum separation in the past between each star in the Greater Pleiades Complex and the core of the Pleiades, using only stars with a non-null \gaia\ radial velocity and RUWE $<1.2$. Stars that do not get closer to the Pleiades core in the past are shown in gray. There is a dynamical age gradient present along the structure, with stars closer to the core appearing lighter and the stars further away appearing darker. This is especially true for UPK 303 and the combination of the HSC 1964 and UPK 545 regions. Gaps or irregularities in the color gradient may reflect uncertainties in the radial velocities rather than physical substructure.}
    \label{fig:expansion}
\end{figure*} 

As open clusters age, two-body interactions preferentially eject lower-mass stars beyond the cluster's tidal radius, where they become subject to Galactic gravitational shear. Stars with slower orbital velocities drift inward toward the Galactic center, speeding ahead to form a leading tidal tail, whereas faster-moving stars migrate outward, slowing down to form a trailing tail \citep{ryden2016dynamics, jerabkova800PcLong2021}.

Under this tidal model, only the stars at the tips of these tails would back-integrate to a minimum separation coinciding with the cluster’s formation epoch. Stars situated closer to the cluster would have evaporated later, yielding dynamical ages younger than those at the tips. Thus, if UPK 303, HSC 1964, and UPK 545 represent tidal tails of the Pleiades, we expect their internal structure to reflect this dynamical age gradient, with stars nearest the core exhibiting minimum separations more recently than those at greater distances.


To trace the dynamical history of the \complex, we performed backward orbit integrations for all stars with non-null \gaia\ radial velocities and RUWE $<$ 1.2. Each trajectory was integrated 120~Myr into the past, along with that of the Pleiades core, and we recorded the epoch of minimum separation between the star and the Pleiades core. Given the limited precision of some \gaia\ radial velocity measurements, individual stellar encounters may be subject to significant uncertainties. We therefore focus on identifying large-scale trends across the structure rather than interpreting the results for any single star.

Our results are shown in Figure~\ref{fig:expansion}. We see the signature of expansion in UPK 303, HSC 1964, and UPK 545. In UPK 303, stars closest to the Pleiades ($\sim$40 pc) have a minimum separation approximately 20--40~Myr ago, whereas the most distant members ($\sim$140 pc) show closest approaches at greater than 100~Myr, matching the canonical age of the Pleiades. Similarly, the extended structure containing HSC 1964, UPK 545, and intermediate stars exhibits the same outward expansion pattern, with minimum separations progressively older at increasing distances. However, beyond $\sim$200 pc from the Pleiades, the dynamical age plateaus at approximately 60--80~Myr. We discuss possible explanations in Section~\ref{subsec:expansion_explanation}.

\subsection{Field Contamination} \label{subsec:contamination}

Given the large spatial extent and low density of this structure, we wanted to estimate how many Pleiades-age stars would be identified by our analysis if their rotation periods followed the distribution observed in the field; that is, the number of field stars expected to survive our rotation period requirement. To this end, we embedded our full membership determination pipeline within a Monte Carlo framework.

We began with the 10,234 stars in our initial target list located within $5~\mathrm{km~s^{-1}}$ of the Pleiades core velocity. Each star was randomly assigned a rotation period drawn from the \kepler\ field star period distribution, weighted by the \kepler\ recovery fraction in 100-K bins centered on the star’s \teff. For each star, we drew a random number between 0 and 1; if the number was below the recovery fraction, a rotation period was assigned, otherwise no period was recorded. Next, we applied an additional completeness correction to account for whether \tess\ would have been able to recover the assigned period. For this, we used the completeness estimates from B25 in 2-day period bins, assuming a Lomb–Scargle (LS) power threshold of 0.05 (following Section~\ref{sec:pleiades}). A second random draw determined whether the assigned period was retained.

We retained only stars with $P_{\rm rot} < 12$ days and passed them through HDBSCAN using the same clustering parameters described in Section~\ref{subsec:hdbscan}. Because many stars were excluded during this process, HDBSCAN did not always recover a structure. We repeated the simulation until HDBSCAN recovered a structure 100 times, which required 717 iterations. This implies a recovery probability of $\sim$13.9\% under the assumption that the true rotation period distribution matches that of the field.

For each of these 100 recovered structures, we computed \mbox{$P(M \mid P_{\mathrm{rot}}, T_{\mathrm{eff}})$} using our Bayesian framework and retained stars with membership probabilities above 50\%. Among the RV-only sample, we found a contamination rate of $3.8 \pm 1.2\%$, indicating $\sim$41 of the 1,075 stars would be captured in our analysis by chance.

We also repeated the full analysis pipeline --- including running our members for the RV-only analysis through \texttt{Comove}, HDBSCAN, and the Bayesian membership inference --- for each of the 100 simulations. In this version of the analysis, the estimated contamination rate was $6.3^{+3.9}_{-4.8}\%$. \added{Given that the majority of the stars in the GPC-1 and GPC-2 regions are identified with the \texttt{comove} step, we believe that the contamination rate is slightly higher in those regions and that this contamination rate should be used for GPC-1 and GPC-2. Contamination is also likely higher in the low-mass stars where slow-down takes longer and hence the rotation diagnostic is less conclusive. Many of the low-mass stars also lack \textit{Gaia} radial velocity measurements and previous studies of contamination rates in clustering analyses show that there tends to be higher contamination when radial velocity measurements are not included \citep{Boyle2023}.} 

\subsection{Binary Contamination}\label{subsec:binaries}

\citet{Simonian2019} found that rapid rotators in the \kepler\ data are primarily tidally synchronized binaries rather than intrinsically young stars. If binaries dominate our sample rather than genuinely rapidly rotating single stars, our conclusions about the size and shape of the complex could be misleading.

To assess potential binary contamination in our catalog, we considered five independent criteria to flag likely binaries from our membership list of 3,091 stars:

\begin{enumerate}
    \item \textit{RUWE} --- Binaries can be identified through an elevated \gaia\ RUWE value \citep{Wood2021}. We flagged stars with a RUWE $>$1.2 (633 stars).
    \item \textit{Position on a CMD} --- A star that appears overluminous on a CMD at a given age is likely a binary. We manually flagged all stars that appear overluminous relative to the Pleiades sequence (533 stars). 
    \item \textit{Fast rotators} --- Stars that are rotating significantly faster than expected for a given age and effective temperature are likely binaries \citep{Bouma2023}. These stars have already been removed from our sample due to how the KDE used to create our sample was defined (see Appendix \ref{sec:full_bayes}).
    \item \gaia's \texttt{non\_single\_star} \textit{flag} --- The \gaia\ \texttt{non\_single\_star} flag indicates if a star has been detected to be an astrometric, spectroscopic, or eclipsing binary by \gaia. We flagged all stars with \texttt{non\_single\_star} set to 1 (55 stars).
    \item \textit{Multiple peaks in the periodogram} --- Binary systems can manifest as additional significant peaks in the periodogram at non-integer multiples of the stellar rotation period. We identified such candidates by flagging secondary signals with power exceeding 50\% of the primary peak and differing from the primary period and its integer harmonics by at least 5\% (414 stars).
\end{enumerate}

The resulting binarity flag is available in Table~\ref{tab:summary_table}.  Applying all of these criteria would reduce our sample from 3,091 to 1,930 stars. The structure of the Greater Pleiades Complex remains consistent after each of these cuts, indicating that the observed structure is indeed composed primarily of young stars rather than binary contaminants.

\section{Discussion}\label{sec:discussion}

\subsection{What is the morphology of the Greater Pleiades Complex?}

Across the six regions considered --- UPK 303, HSC 1964, UPK 545, \upperright, and \lowerright, and AB Dor --- we identify a coherent picture of a spatially extended, dynamically evolving stellar complex. Each region was analyzed using the same multi-pronged methodology, combining spatial and kinematic structure, photometric and rotational age diagnostics, chemical abundances, and orbital traceback modeling.

Three of the regions considered --- UPK 303, HSC 1964, and UPK 545 --- present compelling evidence of physical association with the Pleiades. These populations are spatially detached but linked to the core through filaments of coeval, comoving stars. They exhibit tight clustering in UVW velocity space, photometric and rotational ages that match the core, and metallicities indistinguishable from the Pleiades. Together, these results are highly suggestive that the Pleiades is embedded within a larger, distributed population of coeval stars extending over several hundred parsecs (Figure~\ref{fig:groups}). 


\upperright\ also exhibits many of these features: spatial proximity ($\sim$150 pc), coeval stellar properties, overlapping metallicity distributions and chemical abundances, and broadly similar kinematics. However, the back integration yields a minimum approach of $\sim$90 pc with larger uncertainties, suggesting a possible but less definitive link to the core. The evidence is consistent with this region being more dynamically evolved, externally perturbed, consisting of multiple populations, and/or a single population with significant field contamination. 

AB Dor shows the same features as the \upperright. Its metallicity, chemical abundances, kinematics, and age all directly overlap with the Pleiades. The back integration is again less definitive than for UPK~303, HSC~1964, and UPK~545. We recover 146 stars from the MOCA database membership list here, but we were unable to reproduce previous results showing that AB Dor back integrates to have a minimum separation from the Pleiades around the time of the Pleiades birth \citep{2007MNRAS.377..441O}. Our results are suggestive of a connection with the Pleiades, but are not definitive. 

The membership list of AB Dor has a mixed history. \citet{Barenfeld2013} found only half the core members were chemically homogeneous. \citet{Moranta2022} search for nearby associations recovered only 18 members of AB Dor (as Crius 231), and they were all preferentially in the direction of HSC 1580 / Theia 301. The existing membership of AB Dor may be too heavily contaminated for proper traceback and may explain the larger spread in $UVW$ and the CMD, as well as the large uncertainties in the detailed abundances. AB Dor may also be more an extension of HSC 1580 / Theia 301, rather than being its own, distinct group. Additional membership cleaning using other diagnostics (e.g., lithium) is likely required.

\lowerright\ represents the most ambiguous case. Although it contains a large population of stars with Pleiades-like ages and elemental abundances, the kinematic coherence is weaker --- only a modest overdensity is seen in the VW plane --- and the orbital traceback is less convergent, with a minimum separation of $\sim$100 pc. These factors, coupled with broader scatter in CMD and rotation–temperature space, suggest that this region may be a mixed or marginal population, either loosely related to the Pleiades or dynamically distinct. OSCN~99 is one such population that could be interspersed among a more diffuse population --- a possible connection with the Pleiades warrants followup.

Taken together, these regions trace a complex around the Pleiades, extending hundreds of parsecs beyond the core. The results are consistent with open clusters forming within larger, hierarchically structured associations whose members gradually disperse over time. That three of the regions pass the stringent kinematic, chemical, and temporal consistency tests underscores the importance of looking beyond the tidal radius to recover the full extent of a cluster’s stellar population. The remaining two regions illustrate the challenges of disentangling genuine dispersed members from overlapping field structures, especially when orbital convergence is modest or the local phase space density is low.


\subsection{Tidal tails versus separated groups}\label{subsec:expansion_explanation}


In Section~\ref{subsec:expansion}, we showed that UPK~303, along with the combined structure of HSC~1964 and UPK~545, exhibits a dynamical age gradient along its extent.

For UPK~303, this gradient strongly suggests a tidal origin from the Pleiades cluster. The increasing dynamical age along the tail implies that the inner regions exited the cluster's tidal radius more recently compared to the outer regions. Furthermore, the outermost parts of the tail trace back dynamically to the cluster core at approximately the age of the Pleiades, implying that we have identified the end of the tidal tail.

HSC 1964 and UPK~545 are less likely to be tidal tails. The time of minimum separation rises with increasing distance, as with UPK~303, but it plateaus at $\lesssim80$\,Myr past 150\,pc. This observation suggests two potential scenarios. First, the true endpoint of the trailing tail may extend beyond our observational limits. If the observed structures are purely tidal in origin, UPK~545 would have yielded dynamical ages closer to 120~Myr. The observed plateau at approximately 80~Myr could indicate that the tail indeed extends beyond our current spatial (500~pc) and velocity (5 \kms{}) search limits. Alternatively, external dynamical interactions, such as an encounter with a giant molecular cloud (GMC), might have accelerated the dispersal of the complex. A single close GMC passage can remove roughly 10\% of a cluster's binding energy \citep{ryden2016dynamics}, potentially driving faster dissolution and artificially reducing the inferred dynamical ages.

However, UPK~545 appears more cluster-like, lacking the elongated structure characteristic of tidal evaporation observed in UPK~303. This morphology suggests that UPK~545 formed concurrently with the Pleiades from the same GMC, subsequently drifting apart due to the Galactic tidal field. HSC~1964 might thus represent a composite structure consisting of genuine tidal debris interspersed with stars formed between UPK~545 and the Pleiades core.

Realistically, the observed morphology and dynamics of the Greater Pleiades Complex likely reflect a combination of tidal interactions and external dynamical perturbations. Detailed N-body simulations will be necessary to fully reconstruct and interpret the complex’s spatial and kinematic distribution.

\subsection{Bayesian clustering for identifying diffuse structures}

Most \gaia-only clustering methods rely exclusively on spatial and kinematic information and consequently struggle to identify stellar groups whose density falls well below that of the surrounding field. Because dissolving open clusters rapidly lose spatial coherence, purely kinematic approaches often fail to recover their full extents or reliably link widely separated stellar subgroups into a coherent association. To address this challenge, we developed a Bayesian framework that incorporates stellar rotation periods as an independent, age-sensitive prior. Rotation periods provide an estimate of stellar age through gyrochronology, allowing us to distinguish coeval cluster members from older field contaminants even when spatial overdensities are absent.

Applying this combined rotation–kinematic methodology to the Pleiades cluster significantly expanded our understanding of its physical boundaries and structure. Our approach connected physically separated stellar groups into a single coherent complex, each subgroup independently showing evidence of sharing a common age and origin with the core Pleiades cluster. Importantly, we were able to identify bridges of coeval stars that connect denser subgroups across hundreds of parsecs. These bridges likely comprise stars initially formed in the low-density regions between denser subgroups, later dispersed by tidal forces and shear in the Galactic disk. Without the rotational age prior, it is possible to identify the major structures (e.g., UPK 303, HSC 1964, and UPK 545), but the subtle connections between these groups would remain challenging to separate from the much larger population of field stars. 

Our results thus demonstrate that integrating \tess-derived rotation periods with \gaia-based positions and kinematics is effective in uncovering large-scale stellar associations. This is particularly effective for ages 80-200\,Myr, where most of the bright stars are on the main sequence (so one cannot easily use CMD information), but stars show sufficiently rapid rotation to be identified in \tess. As more progress is made in extracting rotation periods beyond 12\,days from \tess\ \citep{Hattori2025} and with the arrival of \gaia\ epoch photometry it may be possible to extend this to much older populations. Consequently, our framework offers a powerful and scalable tool for mapping out dissolving stellar complexes across the Milky Way disk, significantly enhancing our ability to trace the histories and genealogies of stellar associations.

\section{Conclusion} \label{sec:conclusion}

We developed a Bayesian framework that leverages \tess\ stellar rotation periods and \gaia\ kinematics to calculate membership probabilities for stars in dissolving open clusters. Applying this method to the Pleiades, we identified a large-scale, coeval stellar structure we call the Greater Pleiades Complex, significantly extending the known spatial boundaries of the Pleiades.

The key results from this work are as follows:

\begin{itemize}

\item The Greater Pleiades Complex spans at least 500 pc, containing at least three previously identified subgroups (UPK 303, HSC 1964/Theia 301, and UPK 545/Theia 163) and likely two more (AB Dor and OSCN 99) as well as hundreds of newly recognized co-moving stars outside these groups.

\item All subgroups share consistent ages, elemental abundances, and similar Galactic motions.

\item Orbital back-integrations show that three subgroups approached within $\sim$50 pc of the core cluster $\sim$75 Myr ago, supporting their historical association with the Pleiades.

\item UPK 303's back-integration resembles that of a tidal tail, with the most distant stars giving an age of closest approach closest to that of the Pleiades (120\,Myr). HSC 1964 + UPK 545 resemble a group that broke off from Pleiades $\simeq$75\,Myr.

\item Our Bayesian approach significantly outperforms classical spatial–kinematic clustering by including stellar rotation as an independent age indicator, recovering structures otherwise obscured by field-star contamination.

\item We identified stellar bridges connecting the Pleiades with other young associations, highlighting a more extensive and complex local star formation history.

\end{itemize}

Our methodology is readily scalable to other dissolving stellar associations provided the complex is young enough to exhibit detectable rotation. It complements methods that rely on pre-main sequence stars \citep[e.g.,][]{kerrStarsPhotometricallyYoung2021} that are effective at the youngest ages ($\lesssim$40\,Myr), and searches that use only kinematic and position information \citep[e.g.,][]{kounkelUntanglingGalaxyLocal2019} that can already separate denser groups from the field population. In combination, these provides a powerful approach to unraveling the star formation history of the local Galaxy. We provide our list of Greater Pleiades Complex members and basic information in Table~\ref{tab:example_table}.

\begin{table*}
    \centering
    \caption{Candidate Greater Pleiades Complex Members}
    \begin{tabular}{ccc}
    \hline \hline
        Parameter & Example Value & Description\\
        \hline
        \texttt{dr3\_source\_id} & 63102694601689728 &  Gaia DR3 source identifier \\
        \texttt{TICID} & 61144513 &  Tess Input Catalog identifier\\
        \texttt{ra} & 56.765924 &  Gaia DR3 Right Ascension [deg.]\\
        \texttt{dec} & 20.195681 &  Gaia DR3 Declination [deg.]\\
        \texttt{teff} & 3848.58 & Calculated effective temperature [K]\\
        \texttt{phot\_g\_mean\_mag\_0} & 8.117 & Absolute $G$-band magnitude, corrected for extinction\\
        \texttt{BpmRp0} & 1.887 & Gaia $G_{\rm BP} - G_{\rm RP}$ color, corrected for extinction\\
        \texttt{period} & 8.88 & Measured rotation period [days]\\
        \texttt{period\_unc} & 0.46& Uncertainty on rotation period [days]\\
        \texttt{power} & 0.85& Lomb-Scargle power from rotation period measurement\\
        \texttt{prob} & 0.77 & Bayesian membership probability\\
        \texttt{x} & -8236.7 & Galactocentric $X$-position [pc]\\
        \texttt{y} & 21.2 & Galactocentric $Y$-position [pc]\\
        \texttt{z} & -36.7 & Galactocentric $Z$-position [pc]\\
        \texttt{flag\_pleiades\_cmd} & True & Flag indicating if the star overlaps with the Pleiades on a CMD.\\
        \texttt{flag\_possible\_binary} & False & Flag indicating if the star is in any of the binarity filters in Section~\ref{subsec:binaries}.\\
    \hline
    \multicolumn{3}{c}{\footnotesize \textbf{Note.} This table is published in its entirety in machine-readable format. One entry is shown for guidance regarding form and content.}
    \end{tabular}
    \label{tab:example_table}
\end{table*}


\subsection{Limitations and Future Work} \label{subsec:limitations}

Further extensions of the Pleiades may still remain to be found. With light curves from \gaia\ DR4 and the Rubin Observatory arriving soon, future studies will likely be able to probe more distant regions to search for additional extensions to the Greater Pleiades Complex. Large spectroscopic surveys like the Milky Way Mapper \citep{Meszaros2025} will also improve the availability of radial velocities and provide abundances for more stars. This should provide a stronger baseline to clean the samples. Additional age diagnostics, like lithium, may also improve the lists, which can in turn be used to improve the kinematic back integrations (especially for AB Dor, \lowerright{}, and \upperright{}). \added{As an additional experiment, one could replace the smooth potentials considered here with a clumpy, time-varying potential with various initial conditions in the parent cloud to try to recreate the structure of the Greater Pleiades Complex from a simulation}. 

Another way to improve our membership inference would be to fold each star’s location in the color–magnitude diagram (CMD) into the Bayesian likelihood. Gyrochronology excels for FGK stars --- their tight rotation–age relation cleanly distinguishes them from field interlopers at high temperatures. Stars between 4000 and 5000\,K also benefit from having periods longer than what is seen for most tight binaries \citep{Simonian2019}. At cooler temperatures, the slow‐sequence begins to blur and rotation periods by themselves permit greater contamination. M dwarfs at $\simeq$100\,Myr occupy regions of the CMD that are sparsely populated by single field stars. Interlopers would be limited to equal-mass binaries and/or the most metal-rich M dwarfs \citep{Mann2013a}. We use the tightness and high position of our CMDs (Figure~\ref{fig:upk_303}-\ref{fig:abdor}) as a check, but a CMD-based likelihood term alongside our rotation-period term, one could produce even cleaner, more reliable membership lists. As an example, some cooler stars from Chameleon \citep[$t\lesssim17$~Myr;][]{2023ApJ...954..134K} and Tuc-Hor \citep[$t\approx40$~Myr;][]{Kraus2014} appear in our membership list due to this blurring of the rotation sequence at young ages. These young contaminants can be cleaned from the membership list by removing stars that lie above the Pleiades sequence in the CMD. For now, we have included a flag in Table \ref{tab:example_table} (\texttt{flag\_pleiades\_cmd}) that indicates if a star overlaps with the Pleiades on a CMD. This allows any user to quickly and easily clean our list of likely young star contaminants.

\begin{acknowledgments}
This material is based upon work supported by the National Science Foundation Graduate Research Fellowship Program under Grant No. DGE-2439854. Any opinions, findings, and conclusions or recommendations expressed in this material are those of the authors and do not necessarily reflect the views of the National Science Foundation. This research made use of the Montreal Open Clusters and Associations (MOCA) database, operated at the Montr\'eal Plan\'etarium (J. Gagn\'e et al., in preparation). Andrew W. Boyle thanks the LSST-DA Data Science Fellowship Program, which is funded by LSST-DA, the Brinson Foundation, the WoodNext Foundation, and the Research Corporation for Science Advancement Foundation; his participation in the program has benefited this work. 

AWB and AWM were supported by a grant from NASA's Astrophysics Data Analysis Program (ADAP award 80NSSC24K0619). AWM was also supported by a grant from NASA’s exoplanet research program (XRP 80NSSC25K7148) and the NSF CAREER program (AST-2143763).  LGB  gratefully acknowledges support from the Carnegie Fellowship.

This paper includes data collected by the TESS mission. Funding for the TESS mission is provided by the NASA's Science Mission Directorate. The TESS data used in this paper can be found at MAST \citep{https://doi.org/10.17909/0cp4-2j79}.

This work has made use of data from the European Space Agency (ESA) mission
{\it Gaia} (\url{https://www.cosmos.esa.int/gaia}), processed by the {\it Gaia}
Data Processing and Analysis Consortium (DPAC,
\url{https://www.cosmos.esa.int/web/gaia/dpac/consortium}). Funding for the DPAC
has been provided by national institutions, in particular the institutions
participating in the {\it Gaia} Multilateral Agreement.

Funding for the Sloan Digital Sky Survey V has been provided by the Alfred P. Sloan Foundation, the Heising-Simons Foundation, the National Science Foundation, and the Participating Institutions. SDSS acknowledges support and resources from the Center for High-Performance Computing at the University of Utah. SDSS telescopes are located at Apache Point Observatory, funded by the Astrophysical Research Consortium and operated by New Mexico State University, and at Las Campanas Observatory, operated by the Carnegie Institution for Science. The SDSS web site is \url{www.sdss.org}.

SDSS is managed by the Astrophysical Research Consortium for the Participating Institutions of the SDSS Collaboration, including the Carnegie Institution for Science, Chilean National Time Allocation Committee (CNTAC) ratified researchers, Caltech, the Gotham Participation Group, Harvard University, Heidelberg University, The Flatiron Institute, The Johns Hopkins University, L'Ecole polytechnique f\'{e}d\'{e}rale de Lausanne (EPFL), Leibniz-Institut f\"{u}r Astrophysik Potsdam (AIP), Max-Planck-Institut f\"{u}r Astronomie (MPIA Heidelberg), Max-Planck-Institut f\"{u}r Extraterrestrische Physik (MPE), Nanjing University, National Astronomical Observatories of China (NAOC), New Mexico State University, The Ohio State University, Pennsylvania State University, Smithsonian Astrophysical Observatory, Space Telescope Science Institute (STScI), the Stellar Astrophysics Participation Group, Universidad Nacional Aut\'{o}noma de M\'{e}xico, University of Arizona, University of Colorado Boulder, University of Illinois at Urbana-Champaign, University of Toronto, University of Utah, University of Virginia, Yale University, and Yunnan University.  

Guoshoujing Telescope (the Large Sky Area Multi-Object Fiber Spectroscopic Telescope LAMOST) is a National Major Scientific Project built by the Chinese Academy of Sciences. Funding for the project has been provided by the National Development and Reform Commission. LAMOST is operated and managed by the National Astronomical Observatories, Chinese Academy of Sciences.

\end{acknowledgments}

\begin{contribution}

Per https://credit.niso.org: Conceptualization: AWB and AWM. Data curation: AWB. Formal analysis: AWB, AWM. Funding acquisition: all authors. Investigation: AWB. Methodology: all authors. Project administration: AWB. Resources: AWB, AWM. Software: AWB. Supervision: AWM, LGB. Validation: AWB, LBG. Visualization: AWB, LGB. Writing – original draft: AWB. Writing – review \& editing: all authors.


\end{contribution}

%
\facilities{\tess\ \citep{2015JATIS...1a4003R}, \gaia\ \citep{2016A&A...595A...1G, 2023A&A...674A...1G}, SDSS (MWM) \citep{2025arXiv250706989K}, LAMOST \citep{2012RAA....12.1197C}}

\software{Astropy \citep{2013A&A...558A..33A, 2018AJ....156..123A, 2022ApJ...935..167A},  
tess-point \citep{2020ascl.soft03001B},
matplotlib \citep{Hunter:2007},
pandas \citep{mckinney-proc-scipy-2010, reback2020pandas},
unpopular \citep{hattoriUnpopularPackageDatadriven2021}
          }


\appendix

\section{Bayesian Clustering with Rotation Periods} \label{sec:full_bayes}

Our goal is to calculate the probability that a star is a member of a cluster given a measurement of the star's rotation period. We consider two scenarios. The first is where we recover an estimate of the star's rotation period, where we must consider both the estimated period and the reliability of that estimate. The second is where no period is recovered. A star can still be a cluster member even if its rotation period was not measured (e.g. if the star had a pole-on orientation or a variability amplitude well below the Poisson variability). Thus, we must consider the completeness (the likelihood of measuring a period if the star has a period consistent with membership). For this study, we have defined all rotation measurements with a Lomb-Scargle power $>0.05$ as having a recovered rotation period and those below as a non-recovery. We discuss each scenario in turn below.

\subsection{Successfully Measured Rotation Period} \label{subsec:measured_prot_parameters}

The probability that a star is a cluster member ($M$) given the star's rotation period (\prot) and effective temperature ($T_{\rm eff}$) can be expressed using Bayes' theorem as

\begin{equation} \label{eq:A1}
P\bigl(M \mid P_{\rm rot}, T_{\mathrm{eff}}\bigr)
\;=\;
\frac{P\bigl(P_{\rm rot}\mid M, T_{\mathrm{eff}} \bigr)\,P(M)}
     {P\bigl(P_{\rm rot}\mid M, T_{\mathrm{eff}} \bigr)\,P(M)\;+\;P\bigl(P_{\rm rot}\mid \overline{M}, T_{\mathrm{eff}} \bigr)\,P\bigl(\overline{M}\bigr)}
\end{equation}

where $P(M \mid P_{\rm rot}, T_{\mathrm{eff}})$ is the probability that a star is a member of a cluster given its rotation period and effective temperature, $P(M)$ is the probability of a star being a cluster member, and $P(P_{\rm rot} \mid \overline{M},  T_{\mathrm{eff}})$ is the probability of measuring the rotation period given that the star is not a cluster member. We assumed that if a star is not a cluster member it is a field star, so we can rewrite $P(P_{\rm rot} \mid \overline{M}, T_{\mathrm{eff}})$ as $P(P_{\rm rot} \mid \mathrm{field}, T_{\mathrm{eff}})$. The probability that a star is \textit{not} a cluster member, $P(\overline{M})$, can be re-expressed as $1 - P(M)$. We must additionally weight $P(P_{\rm rot} \mid \rm field, T_{\mathrm{eff}})$ by the chance that a field star will even have a measurable rotation period, which we denote $N_{\rm rot}$.

Next, we take into account the probability that the measured rotation period is correct ($R_{\rm correct}$), a half-period alias of the true rotation period ($R_{\rm alias}$), or entirely incorrect ($R_{\rm incorrect}$). Following B25, an incorrect rotation period is one that is $>3\sigma$ from the true value and $>3\sigma$ from the half-period alias of the true value. We do not take into account the probability that the measured \tess\ rotation period is double the true rotation period as these are exceedingly rare in \tess\ data (see B25). Finally, we also include the likelihood that the star is a member if \tess\ measured the period incorrectly ($P(P_{\rm rot,incorrect} \mid {M})$). All of the probabilities in this paragraph come from from B25. 

Using the above expressions, we expand Equation \ref{eq:A1} to include three terms (one for each possibility):

\begin{equation} \label{eq:A2}
    \begin{aligned} 
    P\bigl(M \mid P_{\rm rot}, T_{\mathrm{eff}} \bigr)
    &=
    \frac{R_{\rm correct}\,P\bigl(P_{\rm rot}\mid M, T_{\mathrm{eff}}\bigr)\,P(M)}
         {P\bigl(P_{\rm rot}\mid M, T_{\mathrm{eff}}\bigr)\,P(M)\;+\;P\bigl(P_{\rm rot}\mid \mathrm{field}, T_{\mathrm{eff}}\bigr)\,\bigl(1 - P(M)\bigr)\,N_{\rm rot}}
    \\
    &+
    \frac{R_{\rm alias}\,P\bigl(2P_{\rm rot}\mid M, T_{\mathrm{eff}}\bigr)\,P(M)}
         {P\bigl(2P_{\rm rot}\mid M, T_{\mathrm{eff}} \bigr)\,P(M)\;+\;P\bigl(2P_{\rm rot}\mid \mathrm{field}, T_{\mathrm{eff}}\bigr)\,\bigl(1 - P(M)\bigr)\,N_{\rm rot}}
    \\
    &+
    \frac{R_{\mathrm{incorrect}}\,P\bigl(P_{\mathrm{rot,incorrect}}\mid M, T_{\mathrm{eff}}\bigr)\,P(M)}
         {P\bigl(P_{\mathrm{rot,incorrect}}\mid M, T_{\mathrm{eff}} \bigr)\,P(M)\;+\;P\bigl(P_{\mathrm{rot,incorrect}}\mid \mathrm{field}, T_{\mathrm{eff}}\bigr)\,\bigl(1 - P(M)\bigr)\,N_{\rm rot}}
    \end{aligned}
\end{equation}

Each term is calculated as follows:

\begin{itemize}
    \item $R_{\rm correct}$ --- \cite{2025arXiv250413262B} provided the framework to calculate the probability that each rotation period measurement is correct, an alias, or entirely incorrect. In this context, ``correct'' refers to a measured period being within $3\sigma$ of the true period.  Briefly summarized, \cite{2025arXiv250413262B} assessed this probability by comparing rotation periods measured by \tess\ to rotation periods for the same stars measured by \ktwo\ to determine where \tess\ succeeds and fails at measuring rotation periods. We use their provided code with \texttt{mode = match} to calculate the probability that each rotation measurement is correct using our measured Lomb-Scargle power and rotation period.
    \item $R_{\rm alias}$ --- Same as for $R_{\rm correct}$, except we use \texttt{mode = alias}.
    \item $R_{\rm incorrect}$ --- Same as for $R_{\rm correct}$, except we use \texttt{mode = not\_recovered}.
    \item $P(M)$ --- HDBSCAN gives a membership probability with its clustering assignments. However, HDBSCAN membership probabilities are usually high, with 6,149 of 6,961 stars in our post-\texttt{Comove} HDBSCAN in Section \ref{subsec:comove} being assigned a membership probability of exactly 1. Looking at equation \ref{eq:A2}, if $P(M) = 1$, all rotation information cancels and we are left with a final $P\bigl(M \mid P_{\rm rot}, T_{\mathrm{eff}} \bigr) = 1$. However, some of the stars that HDBSCAN says have a 100\% chance of being a member can be identified as non-members through their age. For our initial list of HDBSCAN members, we therefore check to see which stars can be identified as non-members via their rotation and position on a color-magnitude diagram. To get $P(M)$, we divide HDBSCAN membership probabilities into 1\% probability bins and weight the HDBSCAN membership probabilities by the fraction of stars in that bin that are consistent with being Pleiades members on a CMD and rotation-effective temperature diagram. 
    \item $P(P_{\mathrm{rot}} \mid M, T_{\mathrm{eff}})$ --- To estimate the likelihood of a star having a measured rotation period given its temperature and that it is a cluster member, we use observed rotation sequences from coeval open clusters. For the Pleiades, we concatenate membership lists for the Pleiades \citep{rebullROTATIONPLEIADESDATA2016}, Pisces-Eridanus \citep{roserCensusNearbyPiscesEridanus2020}, and Blanco-1 \citep{2020MNRAS.492.1008G}. To calculate the distribution of rotation periods at this age, we calculated the Kernel Density Estimate (KDE) with a Gaussian kernel in 100 K wide bins on a grid of temperatures from 3000 K to 6500 K. We chose 100 K bins because the scatter in our calculated effective temperatures is $\pm50$ K. Formally, the KDE with a Gaussian kernel is defined as:

    \begin{equation} \label{eqn:A3}
        f_M\bigl(p \mid T_{\rm eff}\bigr)
        =
        \frac{1}{N\,h}
        \sum_{j=1}^N
        \frac{1}{\sqrt{2\pi}}
        \exp\!\Biggl[
        -\tfrac{1}{2}
        \Bigl(\tfrac{p - p_j^{(M)}}{h}\Bigr)^2
        \Biggr]
    \end{equation}

    where $N$ is the number of stars in the temperature bin, $p_j^{(M)}$ is the cluster rotation periods in that bin, and $h$ is the bandwidth.

    The likelihood of measuring the given $P_{\rm rot}$ given membership in $M$ is then the integral of the product of the KDE and the measurement‐error probability density function (PDF), which we assume to be a Gaussian:

    \begin{equation} \label{eqn:A4}
        P\bigl(P_{\rm rot}\mid M,\,T_{\rm eff}\bigr)
        =
        \int
        f_M\bigl(p\mid T_{\rm eff}\bigr)\,
        \frac{1}{\sqrt{2\pi}\,\sigma_P}
        \exp\!\Bigl[
        -\frac{(P_{\rm rot}-p)^2}{2\,\sigma_P^2}
        \Bigr]
        \,dp
    \end{equation}

    We must also know the uncertainty on the rotation period, $\sigma_P$, to use Equation \ref{eqn:A4}. We calculate uncertainties on each of our measured rotation periods using Equation 1 from \citet{2025arXiv250413262B}.
    
    In practice, the KDE is implemented using \texttt{KDEpy}\footnote{https://kdepy.readthedocs.io/en/latest/index.html} with bandwidth selection based on the Improved Sheather-Jones algorithm. We then integrate Equation \ref{eqn:A4} over a period grid from 0.2 to 176 days (from the shortest to the longest rotation period in our sample, respectively). 
    
    \item $P(P_{\mathrm{rot}} \mid \mathrm{field}, T_{\mathrm{eff}})$ --- We calculate this value using the same methodology as for $P(P_{\mathrm{rot}} \mid M, T_{\mathrm{eff}})$, except we use the observed rotation-effective temperature distribution of stars in the \kepler\ sample from \cite{mcquillanROTATIONPERIODS342014}. For our calculation, we want to use the sample that provides us with the best estimate of the true rotation period distribution of the field star population. \kepler\ is more complete than \tess\ so it is easier to approximate this distribution from the \kepler\ sample than from \tess. The \kepler\ rotation sample does not extend to temperatures cooler than $T_{\rm eff} \sim 3335$ K, so we supplement the \kepler\ sample by including the sample of Zwicky Transient Facility (ZTF) rotators from \cite{2022AJ....164..251L}, which extend to $T_{\rm eff} \sim 3000$ K.
    
    \item $N_{\rm rot}$ --- We get this number from \cite{mcquillanROTATIONPERIODS342014}, who searched for rotation periods around 133,030 stars in the \kepler\ field of view and recovered rotation periods for 34,030 stars (25.6\%).
    \item $P(P_{\mathrm{rot,incorrect}} \mid M, T_{\mathrm{eff}})$ --- In \cite{2025arXiv250413262B}, we compared \ktwo\ and \tess\ periods from the same star and mapped out when \tess\ succeeds and fails at measuring a rotation period as a function of rotation period and effective temperature. We use the same procedure as outlined for $P(P_{\mathrm{rot}} \mid M, T_{\mathrm{eff}})$: we turned the distribution of failed rotation period measurements into a KDE, convolved the KDE with a Gaussian centerered on the measured rotation period with width $\sigma_p$, and integrated over the period grid.
\end{itemize}

\subsection{Unsuccessfully Measured Rotation Period} \label{subsec:no_prot_parameters}

\begin{figure*}
    \centering
    \includegraphics[width=0.5\linewidth]{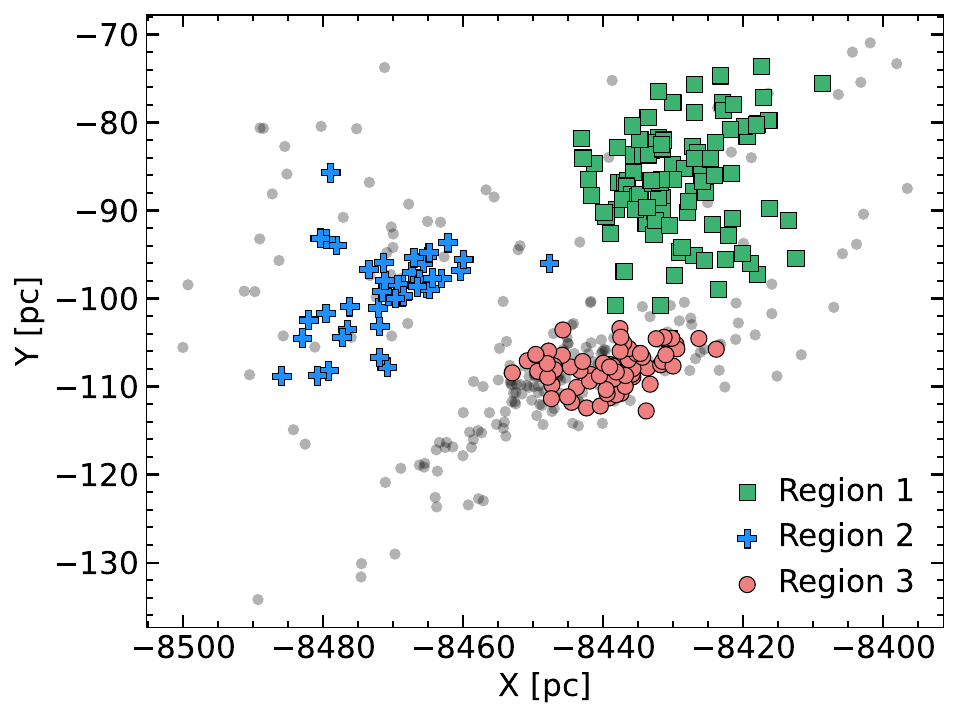}
    \includegraphics[width=1\linewidth]{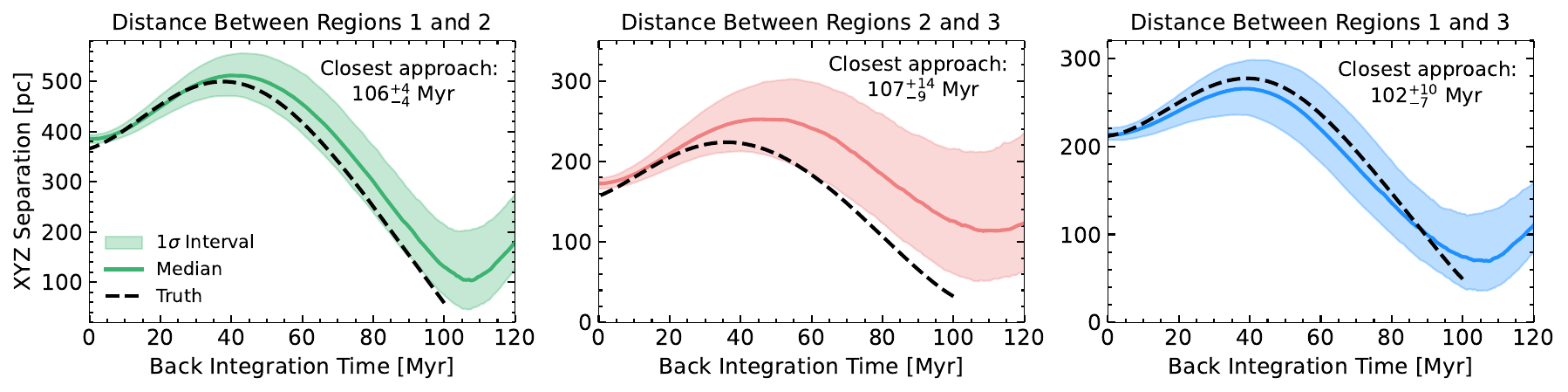}
    \caption{\textit{Top}: The original spatial distributions of Orion members from \citet{2023ApJ...954..134K}. Gray points are outliers that were not considered in our analysis. \textit{Bottom:} Results of our back integration experiment with Orion. The true path the cluster took is shown in black, while the recovered path is shown as a color. The time of closest approach is written on each plot, and is far more precise than the distance at closest approach.}
    \label{fig:orion_back_int}
\end{figure*}

For stars without a successfully measured rotation period, the membership probability is computed via:

\begin{equation} \label{eq:A5}
P\bigl(M \mid P_{\rm rot,failed}\bigr)
\;=\;
\frac{P\bigl(P_{\rm rot,failed}\mid M \bigr)\,P(M)}
     {P\bigl(P_{\rm rot,failed}\mid M \bigr)\,P(M)\;+\;P\bigl(P_{\rm rot,failed}\mid \mathrm{field} \bigr)\,\bigl(1 - P(M)\bigr)}.
\end{equation}

In this case, $P\bigl(P_{\rm rot,failed}\mid M \bigr)$ comes from the completeness from \cite{2025arXiv250413262B}. The completeness ($C$) quantifies the fraction of potentially recoverable rotation periods that were in fact recovered, as a function of Lomb-Scargle power and period. In \cite{2025arXiv250413262B}, we found that completeness is low for high Lomb-Scargle power, indicating that such power cuts eliminate a substantial number of otherwise recoverable periods. For this analysis, we use $1 - C$, which gives the probability that a period was missed due to the power threshold. 

Since we are dealing with stars without measured rotation periods, we must account for the full range of possible true periods that were missed. We compute completeness over the period grid defined in Section~\ref{subsec:measured_prot_parameters}, assuming a LS power $>0.05$ cut. For stars with \prot\ $>40$ days, we set completeness to zero. This gives us completeness as a function of rotation period, $C(P_{\rm rot})$. 

We then marginalize over all possible rotation periods, weighted by the likelihood that a given period was missed:

\begin{equation} \label{eq:A6}
    P\bigl(P_{\mathrm{rot,failed}} \mid M\bigr) = 
    \int_{p_{min}}^{p_{max}} \bigl(1 - C(P_{\mathrm{rot}})\bigr) f_M \bigl(P_{\mathrm{rot}} \mid T_{\mathrm{eff}}\bigr) dP_{\mathrm{rot}}
\end{equation}

\begin{equation} \label{eq:A7}
    P\bigl(P_{\mathrm{rot,failed}} \mid \mathrm{field}\bigr) = 
    N_{\rm rot}\int_{p_{min}}^{p_{max}} \bigl(1 - C(P_{\mathrm{rot}})\bigr) f_{\mathrm{field}} \bigl(P_{\mathrm{rot}} \mid T_{\mathrm{eff}}\bigr) dP_{\mathrm{rot}} + \bigl(1 - N_{\rm rot} \bigr)
\end{equation}

In these expressions, $f_M$ and $f_{\rm field}$ are the KDEs of the cluster and field star period distributions at a given effective temperature, respectively. Equation \ref{eq:A7} contains an additional $1 - N_{\rm rot}$ term because we are assuming that since \cite{mcquillanROTATIONPERIODS342014} was not able to measure rotation periods for $\sim75\%$ of stars, we will not be able to measure rotation periods for those stars, so the probability that we failed to measured the rotation period given that the star is a field member should always be at least $\sim75\%$. The first term in Equation \ref{eq:A7} is then the fraction of field stars with a potentially recoverable rotation period that we missed with \tess. In this way, $P\bigl(P_{\mathrm{rot,failed}} \mid \mathrm{field}\bigr)$ is always bounded between 0.75 and 1. Every other parameter is calculated in the same way as in Section \ref{subsec:measured_prot_parameters}.

\section{On the validity of back integrations with \gaia\ data} \label{sec:back_int_experiment}

In Section \ref{sec:results}, we showed that backwards integrations of UPK 303, HSC 1964, and UPK 545 all reach their closest approach to the Pleiades core at $t\approx75$ Myr ago, despite the cluster’s commonly reported age of $\sim$120 Myr. To determine whether this $\sim$45 Myr discrepancy can be explained by \gaia\ astrometric and kinematic uncertainties, we carried out an idealized Monte Carlo experiment using Orion as a control.

Our procedure comprises three steps:

\begin{enumerate}
    \item \textbf{Forward integration of a coeval control sample}

    We adopt three Orion subgroups (Regions 1–3) from \citet{2023ApJ...954..134K}, which span distinct spatial locations yet share a common age. Their present‐day six‐dimensional positions and velocities are taken from APOGEE DR17 (where available) or \gaia\ DR3. To suppress outliers, each star’s radial velocity is drawn from a normal distribution centered on the subgroup median with a width given by the median absolute deviation. These stars are then integrated forward by 100 Myr in the Milky Way’s potential (\texttt{MilkyWayPotential2022} in \texttt{gala}). The initial spatial configuration is shown in the top panel of Figure \ref{fig:orion_back_int}.

    \item \textbf{Perturbation by \gaia‐like uncertainties}
    
    At $t=100$ Myr, we perturb each star’s proper motion, parallax, and radial velocity by drawing Gaussian offsets whose standard deviations equal the \gaia\ uncertainties of that star. We characterize the DR3 radial velocity uncertainties by fitting the radial velocity uncertainties of the $\sim$34 million stars from \gaia\ DR3 with non‐null radial velocities as a function of their $G_{\rm RVS}$ magnitude, finding median errors of 1.3~km~s$^{-1}$ at $G_{\rm RVS}=12$ and 6.6~km~s$^{-1}$ at $G_{\rm RVS}=14$. For comparison, the formal radial velocity uncertainty is 1.3~km~s$^{-1}$ at $G_{\rm RVS} = 12$ and 6.4~km~s$^{-1}$ at $G_{\rm RVS} = 14$ \citep{2023A&A...674A...5K}. This step replicates the effective error budget affecting our backwards‐integration analysis.

    \item \textbf{Backwards integration}

    For each of 5,000 Monte Carlo realizations, we compute the perturbed median phase‐space coordinate of each subgroup and integrate these medians backwards for 120 Myr. We record the time and separation of closest approach for each pair of regions. The bottom panel of Figure~\ref{fig:orion_back_int} displays the results (solid colored lines and shaded 68\% intervals) alongside the true separations from the forward integration (dashed black lines).
    
\end{enumerate}

The results demonstrate that the time of closest approach is accurately recovered near 100 Myr for all pairs (Regions 1–2: $106^{+4}_{-4}$ Myr; Regions 2–3: $107^{+14}_{-9}$ Myr; Regions 1–3: $102^{+10}_{-7}$ Myr), in close agreement with the true values. However, the distance at closest approach is systematically overestimated: true separations of 59 pc, 32 pc, and 49 pc become recovered medians of $103^{+101}_{-56}$ pc, $113^{+98}_{-61}$ pc, and $70^{+57}_{-33}$ pc, respectively. The large fractional uncertainties on distance --- comparable to their median values --- contrast with the much tighter timing constraints.

This experiment shows that \gaia‐level uncertainties bias our backwards‐integration distances upward while largely preserving the epoch of minimum separation. Although this is an ideal test (we ignore gravitational interactions between stars, dynamical passes with other clusters or molecular clouds, and know a priori which star belongs to which original group), it suggests that current \gaia\ data are sufficient to recover reliable encounter times for a $\sim$100 Myr back-integration but remain limited in their effectiveness for determining how close dispersed stellar groups were when they interacted in the past.

\begin{figure}
    \centering
    \includegraphics[width=1\linewidth]{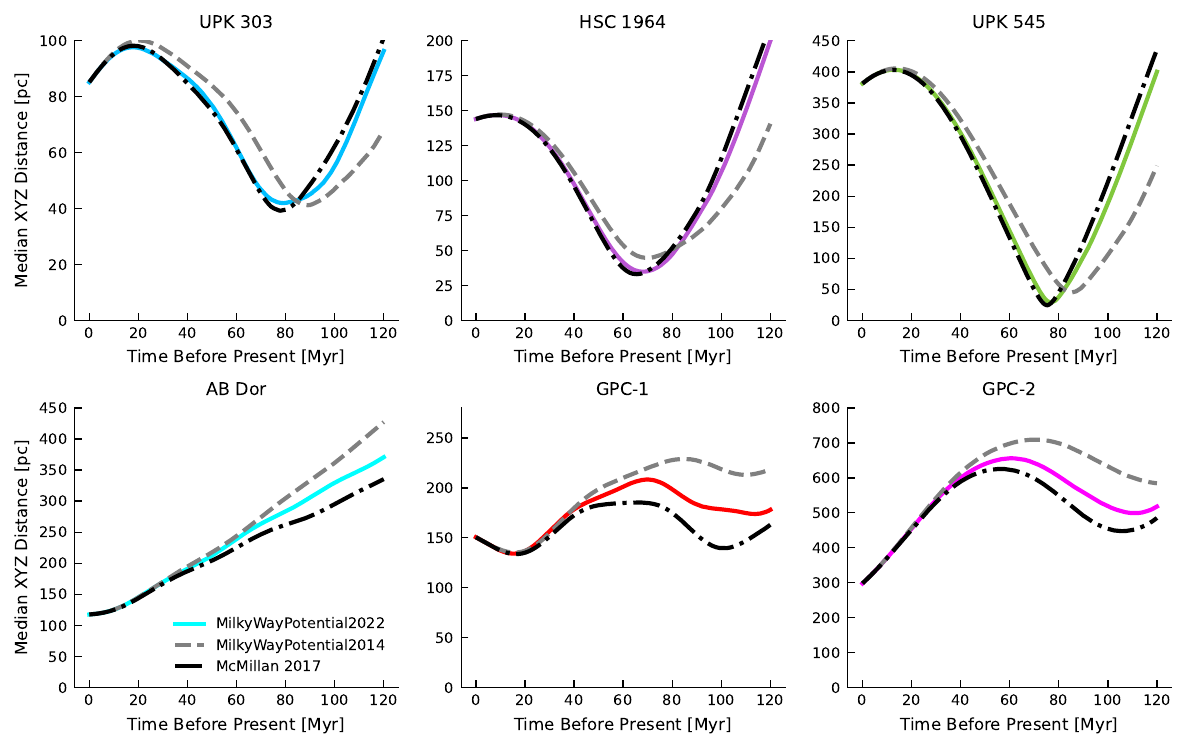}
    \caption{\added{Back integrations of for the six regions of interest relative to the Pleiades core using three different Galactic potential models: \texttt{MilkyWayPotential2022} (solid), \texttt{MWPotential2014} (dashed), and \texttt{McMillan17} (dashed and dotted). Each panel shows the median 3D separation between the region and the Pleiades core as a function of time before the present, integrated to 120 Myr in the past. The three regions with convergent behavior (UPK~303, HSC~1964, and UPK~545) display minima in separation across all three potentials, indicating robustness to model choice. For the remaining regions (AB~Dor, GPC-1, and GPC-2), the trajectories remain inconclusive regardless of potential.}}
    \label{fig:diff_potentials}
\end{figure}

\added{\subsection{Back Integrations with Different Milky Way Potentials}

In Section~\ref{subsec:validation}, we used \texttt{Gala}'s \texttt{MilkyWayPotential2022} to perform back integrations for each of the six regions we studied. As a robustness check, we repeated these integrations using two alternative Galactic potential models implemented in \texttt{galpy}\footnote{\url{http://github.com/jobovy/galpy}}: the \texttt{MWPotential2014} model from \citet{2015ApJS..216...29B} and the \texttt{McMillan17} potential from \citet{2017MNRAS.465...76M}.

The \texttt{MWPotential2014} model includes a power-law bulge with an exponential cutoff, a Miyamoto-Nagai disk, and a dark matter halo described by a Navarro-Frenk-White (NFW) profile. It was empirically calibrated to fit Milky Way kinematics and rotation curves. The \texttt{McMillan17} potential comprises six components: a bulge, thin and thick stellar disks, HI and molecular gas disks, and a dark matter halo, also modeled with an NFW profile. Notably, this model incorporates gas disks as vertically thin structures in the Galactic plane, which can affect stellar trajectories, particularly for populations in the solar neighborhood near the mid-plane.

Figure~\ref{fig:diff_potentials} shows the results of these integrations across the three adopted potentials. For UPK~303, HSC~1964, and UPK~545, all three models yield consistent qualitative behavior: each region approached the Pleiades in the past and followed broadly similar trajectories. For these regions, the differences in the time and distance of closest approach were modest. For instance, UPK~303 reached minimum separations of 42, 41, and 39~pc at 79, 89, and 78~Myr ago under the \texttt{MilkyWayPotential2022}, \texttt{MWPotential2014}, and \texttt{McMillan17} models, respectively. Similar consistency was found for HSC1964, which reached minimum separations of 35, 45, and 33~pc at 68, 70, and 65~Myr ago, respectively; and for UPK~545, with minimum separations of 29, 45, and 25~pc occurring at 77, 85, and 75~Myr ago under the \texttt{MilkyWayPotential2022}, \texttt{MWPotential2014}, and \texttt{McMillan17} models, respectively.

In contrast, the back integrations for GPC-1, GPC-2, and AB Dor yielded inconclusive results across all three potentials. Their separation histories showed no clear minimum or convergence toward the Pleiades, and the trajectories diverged at early times. Nonetheless, the overall trends remained consistent with the inconclusive behavior seen in our baseline analysis.

Taken together, these results indicate that our primary conclusions are not sensitive to reasonable variations in the Galactic potential model. While the absolute timing and distances of closest approach vary slightly across models, our conclusions remain unchanged. A worthy addition to this study would be to consider how young stellar associations evolve in a more complicated galactic potential (e.g. a clumpy galactic potential where the presence of GMCs is considered).}

\section{The Pleiades from Chapel Hill: A Naked-eye Perspective} \label{sec:pleiades_view}

\begin{figure*}
\centering
\includegraphics[width=1\linewidth]{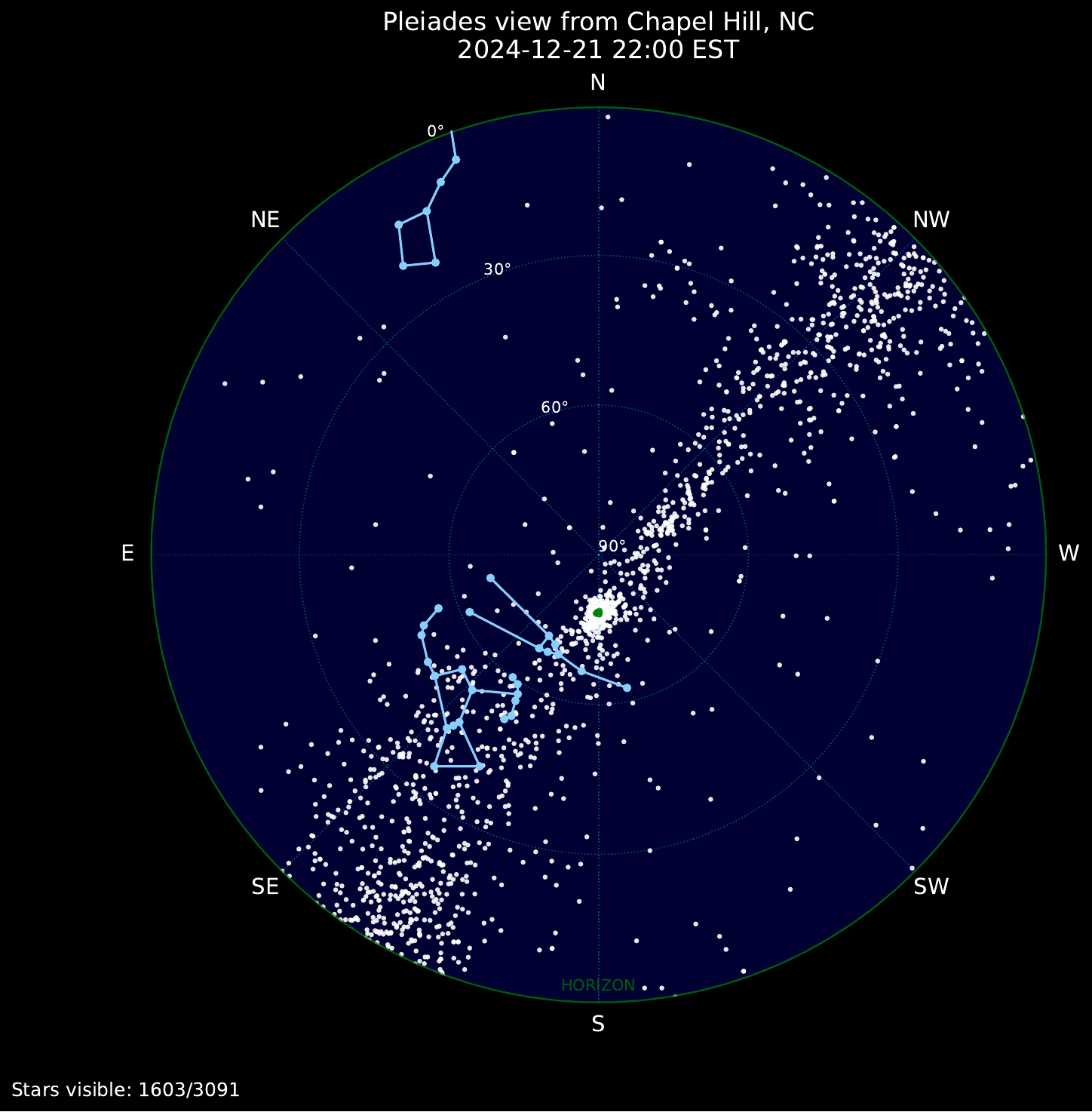}
\caption{The full extent of the Greater Pleiades Complex as it would appear on the night sky if all stars from our membership list were visible to the naked eye. The seven brightest stars in the Pleiades are shown in green, while the stars from our membership list are shown in white. The Big Dipper, Orion, and Taurus are overlaid in blue. Of the 3,091 stars in our membership catalog, 1,603 are above the horizon at the selected observing conditions. The plot is oriented to match what an observer looking to the South would see.}
\label{fig:pleiades_full}
\end{figure*}

From Chapel Hill, North Carolina --- the location of the lead author's home institution --- the Pleiades reaches its highest point in the night sky during December. Figure~\ref{fig:pleiades_full} shows the familiar pattern of the seven brightest stars in this cluster, commonly referred to as the ``Seven Sisters,'' as they appear to the naked eye to an observer in Chapel Hill in December. The full extent of the Greater Pleiades Complex is overlaid behind the Seven Sisters, showing what the Pleiades would look like if all stars from our membership list were visible to the naked eye. The complete distribution spans the sky from horizon to horizon.


\clearpage

\bibliography{PAPER-Prot_Bayes, mannbib, sample7}{}
\bibliographystyle{aasjournalv7}



\end{document}